\documentclass[a4paper,UKenglish]{lipics-v2019}

\usepackage{microtype}

\usepackage[toc,page]{appendix}

\usepackage{booktabs}

\usepackage[utf8]{inputenc}

\usepackage{xcolor}
\definecolor{darkgray}{gray}{0.3}
\usepackage{tikz}
\usetikzlibrary{arrows,shapes,trees, arrows.meta}

\usepackage[plain]{algorithm} 
\usepackage[noend]{algpseudocode}

\hideLIPIcs
\nolinenumbers

\title{Stabilization Time in Weighted Minority Processes}

\author{Pál András Papp}{ETH Zürich, Switzerland}{apapp@ethz.ch}{}{}
\author{Roger Wattenhofer}{ETH Zürich, Switzerland}{wattenhofer@ethz.ch}{}{}

\authorrunning{P.A. Papp and R. Wattenhofer}

\Copyright{Pál András Papp and Roger Wattenhofer}



\relatedversion{The short version of the paper is presented in the 36th Symposium on Theoretical Aspects of Computer Science (STACS'19), 2019.}

\begin{document}

\maketitle

\begin{abstract}
A minority process in a weighted graph is a dynamically changing coloring. Each node repeatedly changes its color in order to minimize the sum of weighted conflicts with its neighbors. We study the number of steps until such a process stabilizes. Our main contribution is an exponential lower bound on stabilization time. We first present a construction showing this bound in the adversarial sequential model, and then we show how to extend the construction to establish the same bound in the benevolent sequential model, as well as in any reasonable concurrent model. Furthermore, we show that the stabilization time of our construction remains exponential even for very strict switching conditions, namely, if a node only changes color when almost all (i.e., any specific fraction) of its neighbors have the same color. Our lower bound works in a wide range of settings, both for node-weighted and edge-weighted graphs, or if we restrict minority processes to the class of sparse graphs. 
 \end{abstract}

\newpage

\setcounter{page}{1}

\section{Introduction}

Given a simple graph and an initial coloring of its nodes, a minority process is a sequence of states (colorings) such that each state is obtained from the previous state by some of the nodes deciding to change their color. Each node, when it has the opportunity to act, switches to the least frequent color in its neighborhood. This may then prompt other neighbors of the node to switch their color, too, leading to a sequence of steps and a dynamically changing coloring. A state is stable when no node in the graph wants to change its color anymore, and the number of steps until a stable state is reached is known as the stabilization time of the process.

Minority processes have numerous applications in different areas where agents in a system are motivated to anti-coordinate with their neighbors. Assume, for instance, a set of wireless devices, each using a given frequency from a predefined set of frequencies for communication. In order to minimize interference with their neighbors, each device may repeatedly decide to switch to the frequency which is the least used in its neighborhood. In another setting, assume that some companies need to decide which product or commodity to produce, and they repeatedly adjust their strategy to avoid competition with specific other companies (that are e.g. geographically close, or share the same costumer base) \cite{KPRanticoor}. Minority processes also appear in a wide range of other areas, including cellular biology \cite{applic3}, physics \cite{applic1, applic2} and social sciences \cite{applic4}.

It is often quite natural to model such settings not only as graphs, but as weighted graphs, since in many applications, either the nodes or edges of the graphs naturally exhibit some kind of weights that define their importance in the minority setting. For example, when selecting products, some competitors may be larger or more resourceful than others, and thus it is more crucial for their neighbors to differentiate from these specific nodes. In the frequency allocation setting, some nodes may handle much more traffic than others, and thus it is more important to avoid interference with such neighbors. Frequency allocation also provides a natural example for edge weights, since the severity of interference can also depend on the distance between neighboring devices, and thus it might be more imperative for nodes to avoid interference with closer neighbors.

The paper considers minority processes in these weighted cases, when the cost function of a node to minimize is not simply the number of its conflicts, but the sum of these conflicts multiplied by the weight of the neighboring node or by the weight of the connecting edge. In such a weighted setting, the only straightforward upper bound on the number of steps is exponential. In this paper, we prove an asymptotically matching lower bound of $2^{\Theta(n)}$, showing that there are weighted graphs where stabilization can indeed last for an exponential number of steps.

For a realistic analysis of stabilization time in applications, some further aspects of the processes are also worth studying. To avoid unreasonably many switches, nodes may decide not to switch color if this benefit is too small. Thus it is often more reasonable to assume a proportional switching rule in the weighted setting, i.e. that a node only decides to change its color if this reduces its cost at least by a given fraction of its weighted degree (or, equivalently, if a large fraction of its neighborhood has the same color). Note that this is a significantly stricter switching rule, and thus proving a lower bound on the number of steps under this rule is a stronger result. Furthermore, in most application areas, the underlying graphs are sparse, i.e. contain only $O(n)$ edges, so it is also interesting to study if the behavior is different when restricting ourselves to sparse graph instances.

There are multiple different models to study minority processes, sequential and concurrent alike. Even in the sequential setting, when only one node switches in each step, we can observe different behaviors depending on the order in which the nodes are selected. For example, this order may be chosen by a benevolent player who aims to minimize stabilization time, or an adversarial player aiming to maximize it. While stabilization time in these models have been studied thoroughly in the related area of majority processes, stabilization time in minority processes has remained open.

In the paper, we present weighted graph constructions that prove an exponential lower bound on stabilization time. Our lower bound holds both for node-weighted and edge-weighted graphs, for any number of colors, and also if we restrict the process to the class of sparse graphs.

The main contributions of the paper are as follows. We first present a construction that shows an exponential lower bound in the adversarial model. Then with further improvements to the construction, we prove that the same bound also holds in the benevolent model. This shows that there are graphs where not only one, but every possible run of the process takes exponential time. Moreover, we also show that the lower bound holds not only for the sequential process, but also in any reasonable concurrent setting. Our lower bounds are shown for a very strict switching rule, when a node is only allowed to switch if a given fraction of its neighbors have the same color. Most surprisingly, our results show that even with this rule, the exponential lower bound holds for any non-trivial fraction of the neighborhood.

\section{Related Work}

The question of stabilization time has only been studied in detail for majority processes. In \cite{majorityW}, the authors devise a weighted graph construction, which exhibits a majority process with $2^{\Theta(n)}$ stabilization time both in the synchronous and the adversarial sequential models (benevolent models are not discussed in this paper). For the unweighted case, the stabilization time of majority processes has been characterized by \cite{majority} in the synchronous, sequential adversarial and sequential benevolent models. The study of \cite{majorityW} also shows further results on some slightly different variants of majority processes in unweigthed graphs. On the other hand, apart from a straightforward $O(n^2)$ upper bound in the unweighted case \cite{hedetniemi, majorityW}, to our knowledge, the stabilization time of minority processes in these models has remained open so far.

However, for unweighted graphs, there are numerous theoretical studies that focus on different properties of stable states, both in case of minority \cite{KPRanticoor, approx0, noUGP, aharoniUGP, UGPrayless} and majority \cite{approx0, SGPclass1, SGPclass2, SGPclass3, SGPclass4, SGPclass5, SGPsurvey} processes.

Minority processes have also been thoroughly studied in special classes of graphs, such as grids, trees or cycles, by the cellular automata community \cite{CA1, CA2, CA3}. However, these results work with unweighted graphs, and a different variant of the minority process which considers the closed neighborhood of nodes. Besides the theoretical results, some of these studies also include an experimental analysis of the process on grids.

Papers working with minority processes almost always consider the basic switching rule, i.e. when nodes switch color for any small amount of improvement (although they sometimes assume different rules for tie-breaking). Some slightly different switching rules, based on distance-2 neighborhood of nodes, are examined in \cite{hedetniemi}; however, the aim of these modified rules is not to achieve earlier stabilization, but to reduce the number of conflicts in the final (stable) state. To our knowledge, however, minority processes have not yet been studied under the proportional switching rule.

\section{Models and Notation}

\paragraph*{Preliminaries and notation.} In the paper, we consider simple, undirected graphs, denoted by $G=(V,E)$, with $V$ being the set of nodes and $E$ the set of edges. The number of nodes is denoted by $n$, the edge between vertices $u$ and $v$ is denoted by $e(u,v)$. In case of \textit{node-weighted graphs}, we assume a positive weight function $w: V \rightarrow {\rm I\!R}^+$ on the nodes of the graph, while for \textit{edge-weighted graphs}, we assume $w: E \rightarrow {\rm I\!R}^+$ on the edges.

For a specific node $v \in V$, we denote by $N(v)$ the neighborhood of $v$. In case of node-weighted graphs, for a set $S \subseteq V$, we denote by $W_{S}$ the sum of weights $\sum_{u \in S} w(u)$. Specifically, we use $W_{N(v)}$ to denote the sum of weights in $v$'s neighborhood.

Given a set of colors $\Gamma$, a \textit{coloring} is a function $C:V \rightarrow \Gamma$. If for some edge $e(u,v)$ we have $C(u)=C(v)$, then we have a \textit{conflict}, and the edge in question is a \textit{conflicting edge}. Generally, the goal of graph coloring is to minimize the number of conflicts in the graph.

We also use the notation $N_S(v) := \{u \: | \: u \in N(v) \text{ and } C(u)=C(v)\}$ and $N_O(v):= N(v) \setminus N_S(v)$ for a node $v$ under a coloring $C$ (the \textit{same-color} and \textit{other-color neighborhood} of $v$, respectively). Note that since we will use these notions in regard to a state of the process (a current coloring of $G$), we assume that the coloring function $C$ is clear from the context, and thus it is not included in the above notation for simplicity.

In both weighted settings, we have a natural cost function $f$ for each node $v$ of the graph. In node-weighted graphs, we define $f(v)= \sum_{u \in N_S(v)} w(u)$, while in the edge-weighted setting, we define the cost function as $f(v)= \sum_{u \in N_S(v)} w(e(u, v))$. The aim of nodes in the minority process is to minimize this cost function. For a color $c \in \Gamma$, let $f_c(v)$ denote the cost that node $v$ would have if it was recolored to color $c$, with the colors of all nodes in $N(v)$ remaining unchanged. Let us denote the preferred color of $v$ by $c^*= \text{arg}\,\text{min}_c \, f_c(v)$; in case of multiple minimal values, we select an arbitrary one of them as $c^*$. When $v$ \textit{switches}, it changes its color to $c^*$. If $f(v)-f_{c^*}(v)$ is above a given threshold, or more generally, if the relation of $f(v)$ and $f_{c^*}(v)$ satisfies a specific condition known as the \textit{switching rule}, then $v$ is \textit{switchable}.

A \textit{minority process} on $G$ is a sequence of colorings $S_0$, $S_1$, ..., known as \textit{states}, where, except for $S_0$, each state $S_i$ can be obtained from $S_{i-1}$ by switching a set of nodes that are switchable in $S_{i-1}$. The state $S_0$ is referred to as the \textit{initial state}. Given a graph and an initial state, the set of nodes to be switched in each step (and thus the entire sequence of states) is determined by the \textit{model}, as discussed below.

We say that a state $S_i$ is \textit{stable} if there are no switchable nodes in $S_i$. A process \textit{stabilizes} if it reaches a stable state; the number of steps until the process stabilizes is the \textit{stabilization time} of the process.

While presenting our construction, we assume node-weighted graphs and $|\Gamma|=2$ available colors. Section \ref{sec:obs} discusses how to generalize our lower bound to edge-weighted graphs or more than 2 colors.

\paragraph*{Models}

We consider minority processes in the following models:

\begin{itemize}
	\item \textbf{Sequential Adversarial (SA):} In each step, only one node switches. This node is chosen by an adversarial player, who aims to maximize the stabilization time.
	
	\item \textbf{Sequential Benevolent (SB):} In each step, only one node switches. This node is chosen by a benevolent player, who aims to minimize the stabilization time.
	
	\item \textbf{Concurrent Benevolent (CB):} In each step, the benevolent player can switch any set of switchable nodes concurrently, in order to minimize the stabilization time.
\end{itemize}

There are many further popular models of minority processes, for example, with synchronous or randomized behavior. However, these models always exhibit a larger stabilization time than model $CB$, since in model $CB$, the benevolent player is free to choose any sequence of (possibly concurrent) steps to minimize stabilization time, and thus he can also simulate the behavior of any of these additional models. Therefore, a lower bound for model $CB$ also implies the same bound in these various other models.

Note that in concurrent models, it is possible that neighboring nodes repeatedly force each other to switch at the same step, cycling through the same colors infinitely. Because of this, related studies in the synchronous model often use an alternative definition of stabilization, also considering a periodically repeating process to be stable. However, the design of our benevolent construction ensures that connected nodes can never be switchable at the same time, and thus in our graphs, even in concurrent models, the process always terminates in a fixed state. Nonetheless, our lower bound also holds with this alternative, more permissive definition of stabilization.

Our lower bound construction for model $SA$ is shown in Section \ref{sec:adv}. Then Section \ref{sec:ben} describes how to extend this construction to the case of model $SB$. Once we present our construction for model $SB$, it will follow that this same construction also proves the lower bound in model $CB$. As the construction heavily restricts the set of selectable sequences, always allowing only a few switchable nodes in the graph, even in model $CB$, the benevolent player has no other option than to execute exactly the same steps as in the sequential case, possibly some of them at the same time. On the other hand, the construction will have specific nodes that alone switch $2^{\Theta(n)}$ times, and thus even with some of the steps executed simultaneously, stabilization takes $2^{\Theta(n)}$ steps.

\paragraph*{Switching rules} Most of the related work studies the following switching rule:

\vspace{7pt}

\noindent \textsc{Rule I} (\textit{Basic Switching}): $v$ is switchable if $W_{N_S(v)} - W_{N_O(v)} > 0$.

\vspace{7pt}

\noindent Here we introduce a stricter switching rule, based on a real parameter $\lambda$ (where $0<\lambda<1$): 

\vspace{7pt}

\noindent \textsc{Rule II} (\textit{Proportional Switching}): $v$ is switchable if $W_{N_S(v)} - W_{N_O(v)} \geq \lambda \cdot W_{N(v)}$.

\vspace{7pt}

\noindent This alternative switching condition is reasonable in many settings where switching comes with a certain cost for the node, and therefore, it is only beneficial when this allows the node to reduce its cost considerably, i.e. by a given factor of $W_{N(v)}$. Since we have $W_{N_S(v)} + W_{N_O(v)} = W_{N(v)}$ in the case of two colors, this condition is equivalent to $W_{N_S(v)} \geq \frac{1+\lambda}{2} \cdot W_{N(v)}$, i.e. that a node is only allowed to switch if $\frac{1+\lambda}{2}$ fraction of its (weighted) neighborhood has the same color. Therefore, if $\lambda$ is close to 1, then Rule II intuitively means that in order to switch $v$ twice, we also have to switch \textit{almost every} neighbor of $v$ in the meantime to make $v$ switchable again for the second time.

While the above definition of Rule II is more intuitive, for the analysis, it is often convenient to express Rule II in another alternative form: $v$ is switchable if $W_{N_S(v)} \geq \Lambda \cdot W_{N_O(v)}$, for some other constant $\Lambda$. One can show that this is equivalent to the definition with a choice of $\Lambda := \frac{1+\lambda}{1-\lambda}$. We will mostly use this alternative $\Lambda$ parameter throughout our analysis.

Our technique proves the lower bound for Rule II with any $\lambda < 1$. However, for ease of presentation, we are first going to describe our construction for a specific parameter value of $\lambda \approx \frac{2}{3}$. Note that $\lambda=\frac{2}{3}$ corresponds to $5$ in the $\Lambda$-notation; let us introduce the new notation $\Lambda_B:=5$ for this base value. We need this extra notation because the construction we present is actually not for $\Lambda=5$, but in fact only for $\Lambda=5-\epsilon$ with any $\epsilon>0$, hence proving the lower bound for Rule II with any $\Lambda<5$ (or, using the $\lambda$-notation, for any $\lambda<\frac{2}{3}$). Note that we have specifically chosen $\lambda > \frac{1}{2}$ for demonstration because some challenges in the construction are easier if $\lambda \leq \frac{1}{2}$.

Given the proof of the lower bound for $\Lambda=5-\epsilon$ with any $\epsilon>0$, we then discuss how to generalize the same construction technique for any other odd integer $\Lambda_B$ as a base value. This proves the lower bound for $\Lambda=7-\epsilon$, $\Lambda=9-\epsilon$, and so on, with any $\epsilon>0$.

Note that $\text{lim}_{\Lambda_B \rightarrow \infty} \lambda=1$, that is, as $\Lambda_B$ goes to infinity, the $\lambda$ value corresponding to $\Lambda_B-\epsilon$ gets arbitrarily close to 1 (this follows from the fact that $\lambda$ can be expressed as $\frac{\Lambda-1}{\Lambda+1}$, by the definition of $\Lambda$). Therefore, we can obtain any $\lambda < 1$ value with an appropriate odd integer $\Lambda_B$ and appropriate $\epsilon$ > 0, and since our construction can be generalized for $\Lambda_B-\epsilon$ with any such $\Lambda_B$ and $\epsilon$, this already establishes the lower bound for every $\lambda \in (0, 1)$.

While it is not required for our lower bound proof, Appendix \ref{App:C} also presents a general method to prove the monotonicity of the lower bound: that is, for any $\lambda_0$ and $\lambda < \lambda_0$ values, given a construction for $\lambda_0$, there is a straightforward way to convert it into a construction for $\lambda$. Note that this monotonicity is trivial in the adversarial case: since any node that is switchable for Rule II with $\lambda_0$ is also switchable for the rule with $\lambda$, the construction for $\lambda_0$ is, without any change, also a valid construction for $\lambda$, exhibiting the same stabilization time. The case is, however, not this simple for benevolent models, where a lower $\lambda$ value may allow a wider set of moves for the benevolent player, which might reduce the stabilization time significantly. Monotonicity in this model can be shown using so-called fixed nodes; see Appendix \ref{App:C} for a discussion.

\paragraph*{Helpful tools and definitions.}

We say that a node $v$ is \textit{dominated} by a subset $S \subseteq N(v)$ if $W_S \geq \frac{\Lambda}{\Lambda+1} \cdot W_{N(v)}$, that is, if $S$ having the same color as $v$ is enough to make $v$ switchable. If $v$ is dominated by a single-node subset $\{u\}$, then we say that $v$ is a \textit{follower node}, and $u$ is the \textit{dominant node} of $v$; this implies that the preferred color of $v$ is always simply the opposite of $u$'s color.

One tool we will frequently use in our constructions is the addition of so-called \textit{fixed node neighbors}. A fixed node is a node that is added to the graph construction in a way that ensures it can never become switchable throughout the process, and thus always keeps its initial color. This can easily be achieved by adding a black and a white \textit{stabilizer node} to the graph, and connecting each fixed node to the stabilizer of the opposite color. If we then assign significantly larger weights to the stabilizer nodes than to all other nodes in the graph (i.e., sufficiently large weights such that each fixed node is a followers of its (opposite-colored) stabilizer node neighbor), then the fixed nodes can indeed never switch throughout the process.

In our construction, each fixed node we add is only connected to one specific node $v$, and its only purpose is to influence the behavior of $v$ in the process (i.e., make it easier or harder to switch $v$ to a specific color). We may add a separate black and a white fixed node neighbor (with any desired weight) to every node $v$ of the construction. However, note that it makes no sense to add more than two fixed neighbors to a node $v$: if we were to add two same-colored fixed neighbors to $v$, we could simply combine the two into one fixed neighbor with the sum of the two weights. Therefore, the use of fixed node neighbors adds at most $2n+2$ extra nodes to the graph, only changing the magnitude of $n$ by a constant factor, and thus it does not affect the exponential nature of stabilization time.

\section{Basic Observations} \label{sec:obs}

\paragraph*{Node or edge weights}

We consider minority processes on both node-weighted and edge-weighted graphs. Note that edge weights have at least as much (in fact, more) expressive power than node weights: assume that we have a graph $G$ with some node weights $w(v)$, and consider the edge-weighted graph that consist of the same nodes and edges, and edge weights are defined as $w(e(u, v)) = w(u) \cdot w(v)$. A minority process in this derived graph behaves the exact same way as in the original, node-weighted graph: for any node $v$, each neighbor $u \in N(v)$ stands for a $\frac{w(u)}{W_{N(v)}}$ portion of $W_{N(v)}$ in the node-weighted case, and $u$ contributes exactly the same $\frac{w(u) \cdot w(v) }{W_{N(v)} \cdot w(v) }$ portion in the derived edge-weighted graph.

This implies that for any node-weighted graph, we can create a corresponding edge-weighted graph with the same stabilization time, regardless of the model. Therefore, when showing the lower bounds of the paper, we only consider node-weighted graph constructions. Our observations imply that the same lower bound will then also hold for edge-weighted graphs.

\paragraph*{Number of colors}

The constructions in the paper assume there are only two available colors: \textit{black} and \textit{white}. However, it is simple to generalize the lower bound to any number of colors. The main idea is to take the lower bound construction for 2 colors, and for each node of the graph and for every additional color, add an extra neighbor with high weight having this color. The process in the resulting graph will behave as if the graph only consisted of the original nodes and the original two colors. A detailed discussion of the technique is available in Appendix \ref{App:C}. The method allows us to generalize the lower bound not only to any constant number of, but also up to $\Theta(n)$ colors.

\paragraph*{Matching upper bound}

While the proof of exponential lower bound is quite involved, it is straightforward to show an exponential upper bound on stabilization time in sequential models. To discuss this upper bound, we briefly return to the case of edge-weighted graphs, as they can exhibit a wider set of behaviors. Since for each node-weighted graph there exists an edge-weighted graph with the same stabilization time, the upper bound on edge-weighted graphs immediately implies the same upper bound on node-weighted graphs.

In an edge-weighted graph, for each state (i.e., coloring of the graph), we can define a \textit{potential} value as the sum of $w(e)$ for all edges $e$ in the graph that are currently conflicting. In sequential models when only one node switches in one step, this potential strictly decreases after every step, since the incentive of the nodes is exactly to reduce the potential in their neighborhood. This allows for a simple upper bound on stabilization time in sequential models: since each state has a fixed potential value and potential is monotonously decreasing throughout the process, each state can be visited at most once. For the case of 2 colors, there are $2^n$ distinct possible states, which implies that stabilization time is upper bounded by $2^n$.

\section{Construction for the Adversarial Case} \label{sec:adv}

We first present a graph construction to show the exponential lower bound in model $SA$.

\begin{theorem}
For Switching Rule II with any $\lambda < 1$, there exists a class of (sparse) weighted graphs with $2^{\Theta(n)}$ stabilization time in model $SA$.
\end{theorem}

While the theorem holds for any $\lambda < 1$, recall that we present the construction for a concrete value of $\lambda \approx \frac{2}{3}$ (that is, $\Lambda=5-\epsilon$ for some small $\epsilon > 0$).

Throughout the presentation of our construction, nodes that are shown vertically higher in figures will always have larger weight than nodes that are placed below. Based on this, we also refer to neighbors of nodes as upper or lower neighbors. We will define the weight of each node in the graph as a function of the weights of the nodes below. As such, one can determine a concrete set of node weights for the construction by following these rules in a bottom-to-top fashion, with the lowermost weights chosen arbitrarily.

The basic idea behind our construction is recursive, and as such, the resulting graph consists of multiple \textit{levels}. Given a construction that exhibits a sequence which switches some specific nodes of the graph $s$ times at least, we show how to extend this graph with a constant number of new nodes (a next level) to obtain another construction where, with the correct choice of sequence, a specific new set of nodes switch $\frac{3}{2} s$ times. With a repeated application of this step, after adding $\ell$ levels, we obtain a set of nodes that switch $\left(\frac{3}{2}\right)^\ell \! \cdot s$ times. Since each new level consists of only $O(1)$ nodes, our graph can contain linearly many levels, yielding a final construction with $2^{\Theta(n)}$ switches.

The key nodes of our graph are the \textit{base nodes}, which appear in 6-tuples with the same weight and same initial color. Each 6-tuple of base nodes has 6 common upper neighbors, known as the \textit{control nodes} for these base nodes, forming a complete bipartite graph. The two 6-tuples together comprise a level of our construction (see Figure \ref{fig:basic}).

The 6 control nodes in a level also all have the same weight; let us denote this weight by $w(v_c)$. The main idea of the construction is to choose $w(v_c)$ sufficiently large such that 5 of the 6 control nodes already dominate each of the base nodes below. Assuming that one of the base node $v_b$ has further (lower) neighbors of weight $w_L$ altogether, this requires $5 \cdot w(v_c) \geq \Lambda \cdot (w(v_c)+w_L)$ to hold, which can be ensured by a choice of $w(v_c) \geq \frac{5-\epsilon}{\epsilon} \cdot w_L$ for our current $\Lambda=5- \epsilon$. Thus we can select sufficiently large weights such that a base node $v_b$ is indeed switchable whenever 5 out of 6 control nodes have the same color as $v_b$.

\begin{figure}
\centering
\minipage{0.42\textwidth}
\centering
	\begin{tikzpicture}

	\draw (27pt,18pt) -- (33pt,24pt);
	\draw (27pt,24pt) -- (33pt,18pt);

	\draw (-10pt,0pt) -- (28.25pt,17.3pt);
	\draw (6pt,0pt) -- (28.95pt,17.3pt);
	\draw (22pt,0pt) -- (29.65pt,17.3pt);
	\draw (38pt,0pt) -- (30.35pt,17.3pt);
	\draw (54pt,0pt) -- (31.05pt,17.3pt);
	\draw (70pt,0pt) -- (31.75pt,17.3pt);
	
	\draw (-10pt,42pt) -- (28.25pt,24.7pt);
	\draw (6pt,42pt) -- (28.95pt,24.7pt);
	\draw (22pt,42pt) -- (29.65pt,24.7pt);
	\draw (38pt,42pt) -- (30.35pt,24.7pt);
	\draw (54pt,42pt) -- (31.05pt,24.7pt);
	\draw (70pt,42pt) -- (31.75pt,24.7pt);
	
	\draw[black, fill=white] (-10pt,42pt) circle (0.9ex);
	\draw[black, fill=white] (6pt,42pt) circle (0.9ex);
	\draw[black, fill=white] (22pt,42pt) circle (0.9ex);
	\draw[black, fill=white] (38pt,42pt) circle (0.9ex);
	\draw[fill=black] (54pt,42pt) circle (0.9ex);
	\draw[black, fill=white] (70pt,42pt) circle (0.9ex);
	
	\draw[fill=black] (-10pt,0pt) circle (0.9ex);
	\draw[fill=black] (6pt,0pt) circle (0.9ex);
	\draw[fill=black] (22pt,0pt) circle (0.9ex);
	\draw[fill=black] (38pt,0pt) circle (0.9ex);
	\draw[fill=black] (54pt,0pt) circle (0.9ex);
	\draw[fill=black] (70pt,0pt) circle (0.9ex);
	
\end{tikzpicture}
	\caption{A 6-tuple of base nodes (below) and control nodes (above). The symbol $\times$ denotes a complete bipartite connection.}
	\label{fig:basic}
	\vspace{40pt}
	\begin{tikzpicture}

	\draw (28pt,18pt) -- (32pt,22pt);
	\draw (28pt,22pt) -- (32pt,18pt);

	\draw (0pt,0pt) -- (28.75pt,17.3pt);
	\draw (12pt,0pt) -- (29.25pt,17.3pt);
	\draw (24pt,0pt) -- (29.75pt,17.3pt);
	\draw (36pt,0pt) -- (30.25pt,17.3pt);
	\draw (48pt,0pt) -- (30.75pt,17.3pt);
	\draw (60pt,0pt) -- (31.25pt,17.3pt);
	
	\draw (0pt,40pt) -- (28.75pt,22.7pt);
	\draw (12pt,40pt) -- (29.25pt,22.7pt);
	\draw (24pt,40pt) -- (29.75pt,22.7pt);
	\draw (36pt,40pt) -- (30.25pt,22.7pt);
	\draw (48pt,40pt) -- (30.75pt,22.7pt);
	\draw (92pt,39pt) -- (31.25pt,22.7pt);
	
	\draw (118pt,18pt) -- (122pt,22pt);
	\draw (118pt,22pt) -- (122pt,18pt);

	\draw (90pt,0pt) -- (118.75pt,17.3pt);
	\draw (102pt,0pt) -- (119.25pt,17.3pt);
	\draw (114pt,0pt) -- (119.75pt,17.3pt);
	\draw (126pt,0pt) -- (120.25pt,17.3pt);
	\draw (138pt,0pt) -- (120.75pt,17.3pt);
	\draw (150pt,0pt) -- (121.25pt,17.3pt);
	
	\draw (58pt,39pt) -- (118.75pt,22.7pt);
	\draw (102pt,40pt) -- (119.25pt,22.7pt);
	\draw (114pt,40pt) -- (119.75pt,22.7pt);
	\draw (126pt,40pt) -- (120.25pt,22.7pt);
	\draw (138pt,40pt) -- (120.75pt,22.7pt);
	\draw (150pt,40pt) -- (121.25pt,22.7pt);
	
	\draw[black, fill=white] (0pt,40pt) circle (0.7ex);
	\draw[black, fill=white] (12pt,40pt) circle (0.7ex);
	\draw[black, fill=white] (24pt,40pt) circle (0.7ex);
	\draw[black, fill=white] (36pt,40pt) circle (0.7ex);
	\draw[black, fill=white] (48pt,40pt) circle (0.7ex);
	\draw[black, fill=white] (60pt,40pt) circle (0.7ex);
	
	\draw[fill=black] (90pt,40pt) circle (0.7ex);
	\draw[fill=black] (102pt,40pt) circle (0.7ex);
	\draw[fill=black] (114pt,40pt) circle (0.7ex);
	\draw[fill=black] (126pt,40pt) circle (0.7ex);
	\draw[fill=black] (138pt,40pt) circle (0.7ex);
	\draw[fill=black] (150pt,40pt) circle (0.7ex);
	
	\draw[fill=black] (0pt,0pt) circle (0.7ex);
	\draw[fill=black] (12pt,0pt) circle (0.7ex);
	\draw[fill=black] (24pt,0pt) circle (0.7ex);
	\draw[fill=black] (36pt,0pt) circle (0.7ex);
	\draw[fill=black] (48pt,0pt) circle (0.7ex);
	\draw[fill=black] (60pt,0pt) circle (0.7ex);
	
	\draw[black, fill=white] (90pt,0pt) circle (0.7ex);
	\draw[black, fill=white] (102pt,0pt) circle (0.7ex);
	\draw[black, fill=white] (114pt,0pt) circle (0.7ex);
	\draw[black, fill=white] (126pt,0pt) circle (0.7ex);
	\draw[black, fill=white] (138pt,0pt) circle (0.7ex);
	\draw[black, fill=white] (150pt,0pt) circle (0.7ex);
	
\end{tikzpicture}
	\caption{Final structure of a level, with two distinct 6-tuples of base and control nodes.}
	\vspace{8pt}
	\label{fig:double}
\endminipage\hfill
\hspace{0.05\textwidth}	
\minipage{0.48\textwidth}
\centering
	\begin{tikzpicture}

	\node[anchor=north] at (30pt,29pt) {\small \textsl{control nodes}};
	\node[anchor=north] at (30pt,40pt) {\small \textsl{Color of}};
	
	\node[anchor=north] at (130pt,29pt) {\small \textsl{afterwards}};
	\node[anchor=north] at (130pt,40pt) {\small \textsl{Color of $v_b$}};

	\draw[black, fill=white] (0pt,0pt) circle (0.7ex);
	\draw[black, fill=white] (12pt,0pt) circle (0.7ex);
	\draw[black, fill=white] (24pt,0pt) circle (0.7ex);
	\draw[black, fill=white] (36pt,0pt) circle (0.7ex);
	\draw[fill=black] (48pt,0pt) circle (0.7ex);
	\draw[black, fill=white] (60pt,0pt) circle (0.7ex);
	
	\draw[arrows=-stealth] (30pt,-9pt) -- (30pt,-17pt);
	
	\draw[densely dotted, semithick] (-2.5pt,-29pt) -- (39pt,-29pt);
	
	\draw[fill=black] (0pt,-24pt) circle (0.7ex);
	\draw[fill=black] (12pt,-24pt) circle (0.7ex);
	\draw[fill=black] (24pt,-24pt) circle (0.7ex);
	\draw[fill=black] (36pt,-24pt) circle (0.7ex);
	\draw[fill=black] (48pt,-24pt) circle (0.7ex);
	\draw[black, fill=white] (60pt,-24pt) circle (0.7ex);
	
	\draw[arrows=-stealth] (30pt,-33pt) -- (30pt,-41pt);
	
	\draw[densely dotted, semithick] (9.5pt,-53pt) -- (51pt,-53pt);
	
	\draw[fill=black] (0pt,-48pt) circle (0.7ex);
	\draw[black, fill=white] (12pt,-48pt) circle (0.7ex);
	\draw[black, fill=white] (24pt,-48pt) circle (0.7ex);
	\draw[black, fill=white] (36pt,-48pt) circle (0.7ex);
	\draw[black, fill=white] (48pt,-48pt) circle (0.7ex);
	\draw[black, fill=white] (60pt,-48pt) circle (0.7ex);
	
	\draw[arrows=-stealth] (30pt,-57pt) -- (30pt,-65pt);
	
	\draw[densely dotted, semithick] (21.5pt,-77pt) -- (63pt,-77pt);
	
	\draw[fill=black] (0pt,-72pt) circle (0.7ex);
	\draw[black, fill=white] (12pt,-72pt) circle (0.7ex);
	\draw[fill=black] (24pt,-72pt) circle (0.7ex);
	\draw[fill=black] (36pt,-72pt) circle (0.7ex);
	\draw[fill=black] (48pt,-72pt) circle (0.7ex);
	\draw[fill=black] (60pt,-72pt) circle (0.7ex);
	
	\draw[arrows=-stealth] (30pt,-81pt) -- (30pt,-89pt);
	
	\draw[densely dotted, semithick] (33.5pt,-101pt) -- (63pt,-101pt);
	\draw[densely dotted, semithick] (-3pt,-101pt) -- (2.5pt,-101pt);
	
	\draw[black, fill=white] (0pt,-96pt) circle (0.7ex);
	\draw[black, fill=white] (12pt,-96pt) circle (0.7ex);
	\draw[fill=black] (24pt,-96pt) circle (0.7ex);
	\draw[black, fill=white] (36pt,-96pt) circle (0.7ex);
	\draw[black, fill=white] (48pt,-96pt) circle (0.7ex);
	\draw[black, fill=white] (60pt,-96pt) circle (0.7ex);
	
	\draw[arrows=-stealth] (30pt,-105pt) -- (30pt,-113pt);
	
	\draw[densely dotted, semithick] (45.5pt,-125pt) -- (63pt,-125pt);
	\draw[densely dotted, semithick] (-3pt,-125pt) -- (14.5pt,-125pt);
	
	\draw[fill=black] (0pt,-120pt) circle (0.7ex);
	\draw[fill=black] (12pt,-120pt) circle (0.7ex);
	\draw[fill=black] (24pt,-120pt) circle (0.7ex);
	\draw[black, fill=white] (36pt,-120pt) circle (0.7ex);
	\draw[fill=black] (48pt,-120pt) circle (0.7ex);
	\draw[fill=black] (60pt,-120pt) circle (0.7ex);
	
	\draw[arrows=-stealth] (30pt,-129pt) -- (30pt,-137pt);
	
	\draw[densely dotted, semithick] (57.5pt,-149pt) -- (63pt,-149pt);
	\draw[densely dotted, semithick] (-3pt,-149pt) -- (26.5pt,-149pt);
	
	\draw[black, fill=white] (0pt,-144pt) circle (0.7ex);
	\draw[black, fill=white] (12pt,-144pt) circle (0.7ex);
	\draw[black, fill=white] (24pt,-144pt) circle (0.7ex);
	\draw[black, fill=white] (36pt,-144pt) circle (0.7ex);
	\draw[fill=black] (48pt,-144pt) circle (0.7ex);
	\draw[black, fill=white] (60pt,-144pt) circle (0.7ex);
	
	
	\draw[fill=black] (130pt,0pt) circle (0.7ex);
	
	\draw[dashed, arrows=-stealth] (90pt,-24pt) -- (110pt,-24pt);
	\draw[arrows=-stealth] (124pt,-20pt) to[out=40,in=140] (136.5pt,-20.5pt);
	\draw[fill=black] (120pt,-24pt) circle (0.7ex);
	\draw[black, fill=white] (140pt,-24pt) circle (0.7ex);
	
	\draw[dashed, arrows=-stealth] (90pt,-48pt) -- (110pt,-48pt);
	\draw[arrows=-stealth] (124pt,-44pt) to[out=40,in=140] (136.5pt,-44.5pt);
	\draw[black, fill=white] (120pt,-48pt) circle (0.7ex);
	\draw[fill=black] (140pt,-48pt) circle (0.7ex);
	
	\draw[dashed, arrows=-stealth] (90pt,-72pt) -- (110pt,-72pt);
	\draw[arrows=-stealth] (124pt,-68pt) to[out=40,in=140] (136.5pt,-68.5pt);
	\draw[fill=black] (120pt,-72pt) circle (0.7ex);
	\draw[black, fill=white] (140pt,-72pt) circle (0.7ex);
	
	\draw[dashed, arrows=-stealth] (90pt,-96pt) -- (110pt,-96pt);
	\draw[arrows=-stealth] (124pt,-92pt) to[out=40,in=140] (136.5pt,-92.5pt);
	\draw[black, fill=white] (120pt,-96pt) circle (0.7ex);
	\draw[fill=black] (140pt,-96pt) circle (0.7ex);

	\draw[dashed, arrows=-stealth] (90pt,-120pt) -- (110pt,-120pt);
	\draw[arrows=-stealth] (124pt,-116pt) to[out=40,in=140] (136.5pt,-116.5pt);
	\draw[fill=black] (120pt,-120pt) circle (0.7ex);
	\draw[black, fill=white] (140pt,-120pt) circle (0.7ex);
	
	\draw[dashed, arrows=-stealth] (90pt,-144pt) -- (110pt,-144pt);
	\draw[arrows=-stealth] (124pt,-140pt) to[out=40,in=140] (136.5pt,-140.5pt);
	\draw[black, fill=white] (120pt,-144pt) circle (0.7ex);
	\draw[fill=black] (140pt,-144pt) circle (0.7ex);

\end{tikzpicture}
	\caption{A control sequence of 6 steps, each time switching a 4-node subset of the control nodes (marked by a dotted line). The resulting switch of the base nodes is shown on the right.}
	\label{fig:combinat}
\endminipage\hfill
\end{figure}

Note that from the initial state shown in Figure \ref{fig:basic}, we only need to switch 4 of the 6 control nodes (from white to black) in order to force a base node $v_b$ below to switch to white. In fact, we can specify a sequence of 4-node subsets of the control nodes such that every time we switch the next subset in the sequence, we once again have 5 control nodes with the same color that $v_b$ currently has, and therefore $v_b$ can be switched again. A possible such sequence is shown in Figure \ref{fig:combinat}; we refer to this as the \textit{control sequence}. The sequence has a couple of convenient properties: each control node is switched exactly 4 times throughout the sequence, and each control node (and also $v_b$) returns to its initial color at the end of the sequence.

This is exactly the technique that allows us to increase the number of switches by a factor of $\frac{3}{2}$ within each level of the construction. If the upper levels provide a way to switch each of the 6 control nodes in the current level $s$ times, then this allows us to execute the control sequence $\frac{s}{4}$ times, and each such execution switches the base nodes in the current level 6 times, adding up to $\frac{6}{4}s$ switches for each of the 6 base nodes.

It only remains to connect the different levels of our recursive construction. It comes as a natural first idea that the 6-tuple of base nodes in a level could also directly take the role of the control nodes in the level below. The first difficulty to overcome with this approach is the color of the nodes in question: while all 6 base nodes of a level have the same color (say, initially black), the control nodes initially have mixed color (5 white and 1 black) in the control sequence. We can overcome this by duplicating the structure in Figure \ref{fig:basic} in the opposite initial color, and redefining a level as these two bipartite graphs together. Since a level now consists of 12 base nodes, 6 white and 6 black initially, we can reorganize these nodes into two appropriate groups (5 white + 1 black, 5 black + 1 white) to act as the control nodes of the next level (see Figure \ref{fig:double}).

There is a further problem with using the base nodes directly as the control nodes of the level below: our level design only provides a way to switch a 6-tuple of base nodes \textit{together} (that is, consecutively in any order). However, in order to execute the control sequence, we need to be able to switch specific subsets of the control nodes. For example, in the sequence of Figure \ref{fig:combinat}, the second node from the left has already switched twice before the rightmost node ever switches. Thus, the fact that we can switch both 6-tuples of base nodes $s$ times does not yet imply that we can switch specific 4-node subsets of them in the given order, as needed for the control sequence.

\begin{figure}
\centering
\minipage{0.31\textwidth}
\centering
	\vspace{10pt}
	 \definecolor{dgray}{gray}{0.4}

\begin{tikzpicture}

	\draw (0pt,0pt) -- (0pt,72pt);

	\draw[fill=black] (0pt,0pt) circle (0.9ex);
	\draw[black, fill=white] (0pt,18pt) circle (0.9ex);
	\draw[fill=black] (0pt,36pt) circle (0.9ex);
	\draw[black, fill=white] (0pt,54pt) circle (0.9ex);
	\draw[fill=black] (0pt,72pt) circle (0.9ex);
	
	\draw[thick, arrows=-stealth] (3pt,77pt) to[out=70,in=110] (20.5pt,76.5pt);
	
	\draw (23pt,0pt) -- (23pt,72pt);

	\draw[fill=black] (23pt,0pt) circle (0.9ex);
	\draw[black, fill=white] (23pt,18pt) circle (0.9ex);
	\draw[fill=black] (23pt,36pt) circle (0.9ex);
	\draw[black, fill=white] (23pt,54pt) circle (0.9ex);
	\draw[black, fill=white] (23pt,72pt) circle (0.9ex);
	
	\draw[densely dotted, thick, dgray] (16pt,78.5pt) -- (16pt,47.5pt) -- (30pt,47.5pt);
	\draw[densely dotted, thick, dgray] (30pt,47.5pt) -- (30pt,78.5pt) -- (16pt,78.5pt);
	
	\draw[thick, arrows=-stealth] (31pt,36pt) -- (39pt,36pt);
	
	\draw (46pt,0pt) -- (46pt,72pt);

	\draw[fill=black] (46pt,0pt) circle (0.9ex);
	\draw[black, fill=white] (46pt,18pt) circle (0.9ex);
	\draw[fill=black] (46pt,36pt) circle (0.9ex);
	\draw[fill=black] (46pt,54pt) circle (0.9ex);
	\draw[black, fill=white] (46pt,72pt) circle (0.9ex);
	
	\draw[densely dotted, thick, dgray] (39pt,60.5pt) -- (39pt,29.5pt) -- (53pt,29.5pt);
	\draw[densely dotted, thick, dgray] (53pt,29.5pt) -- (53pt,60.5pt) -- (39pt,60.5pt);
	
	\draw[thick, arrows=-stealth] (54pt,36pt) -- (62pt,36pt);
	
	\draw (69pt,0pt) -- (69pt,72pt);

	\draw[fill=black] (69pt,0pt) circle (0.9ex);
	\draw[black, fill=white] (69pt,18pt) circle (0.9ex);
	\draw[black, fill=white] (69pt,36pt) circle (0.9ex);
	\draw[fill=black] (69pt,54pt) circle (0.9ex);
	\draw[black, fill=white] (69pt,72pt) circle (0.9ex);
	
	\draw[densely dotted, thick, dgray] (62pt,42.5pt) -- (62pt,11.5pt) -- (76pt,11.5pt);
	\draw[densely dotted, thick, dgray] (76pt,11.5pt) -- (76pt,42.5pt) -- (62pt,42.5pt);
	
	\draw[thick, arrows=-stealth] (77pt,36pt) -- (85pt,36pt);
	
	\draw (92pt,0pt) -- (92pt,72pt);

	\draw[fill=black] (92pt,0pt) circle (0.9ex);
	\draw[fill=black] (92pt,18pt) circle (0.9ex);
	\draw[black, fill=white] (92pt,36pt) circle (0.9ex);
	\draw[fill=black] (92pt,54pt) circle (0.9ex);
	\draw[black, fill=white] (92pt,72pt) circle (0.9ex);
	
	\draw[densely dotted, thick, dgray] (85pt,24.5pt) -- (85pt,-6.5pt) -- (99pt,-6.5pt);
	\draw[densely dotted, thick, dgray] (99pt,-6.5pt) -- (99pt,24.5pt) -- (85pt,24.5pt);

\end{tikzpicture}
	\vspace{15.5pt}
	\caption{When a conflict is created at the top of the chain, then switching the nodes one by one propagates this conflict down through the chain.}
	\label{fig:prop}
	\vspace{9.5pt}
\endminipage\hfill
\hspace{0.01\textwidth}	
\minipage{0.61\textwidth}
\centering
		\begin{subfigure}[b]{0.406\textwidth}
			\centering
			\definecolor{dgray}{gray}{0.4}

\begin{tikzpicture}

	\draw (0pt,0pt) -- (0pt,72pt);

	\draw[fill=black] (0pt,0pt) circle (0.9ex);
	\draw[black, fill=white] (0pt,18pt) circle (0.9ex);
	\draw[fill=black] (0pt,36pt) circle (0.9ex);
	\draw[black, fill=white] (0pt,54pt) circle (0.9ex);
	\draw[fill=black] (0pt,72pt) circle (0.9ex);
	
	\draw[thick, arrows=-stealth] (3pt,77pt) to[out=70,in=110] (20.5pt,76.5pt);
	
	\draw (23pt,0pt) -- (23pt,72pt);

	\draw[fill=black] (23pt,0pt) circle (0.9ex);
	\draw[fill=black] (23pt,18pt) circle (0.9ex);
	\draw[black, fill=white] (23pt,36pt) circle (0.9ex);
	\draw[fill=black] (23pt,54pt) circle (0.9ex);
	\draw[black, fill=white] (23pt,72pt) circle (0.9ex);
	
	\draw[densely dotted, thick, dgray] (16pt,24.5pt) -- (16pt,-6.5pt) -- (30pt,-6.5pt);
	\draw[densely dotted, thick, dgray] (30pt,-6.5pt) -- (30pt,24.5pt) -- (16pt,24.5pt);
	
	\draw[thick, arrows=-stealth] (26pt,77pt) to[out=70,in=110] (43.5pt,76.5pt);
	
	\draw (46pt,0pt) -- (46pt,72pt);

	\draw[fill=black] (46pt,0pt) circle (0.9ex);
	\draw[fill=black] (46pt,18pt) circle (0.9ex);
	\draw[fill=black] (46pt,36pt) circle (0.9ex);
	\draw[black, fill=white] (46pt,54pt) circle (0.9ex);
	\draw[fill=black] (46pt,72pt) circle (0.9ex);
	
	\draw[densely dotted, thick, dgray] (39pt,42.5pt) -- (39pt,11.5pt) -- (53pt,11.5pt);
	\draw[densely dotted, thick, dgray] (53pt,11.5pt) -- (53pt,42.5pt) -- (39pt,42.5pt);
	
	\draw[thick, arrows=-stealth] (49pt,77pt) to[out=70,in=110] (66.5pt,76.5pt);
	
	\draw (69pt,0pt) -- (69pt,72pt);

	\draw[fill=black] (69pt,0pt) circle (0.9ex);
	\draw[fill=black] (69pt,18pt) circle (0.9ex);
	\draw[fill=black] (69pt,36pt) circle (0.9ex);
	\draw[fill=black] (69pt,54pt) circle (0.9ex);
	\draw[black, fill=white] (69pt,72pt) circle (0.9ex);
	
	\draw[densely dotted, thick, dgray] (62pt,60.5pt) -- (62pt,29.5pt) -- (76pt,29.5pt);
	\draw[densely dotted, thick, dgray] (76pt,29.5pt) -- (76pt,60.5pt) -- (62pt,60.5pt);
	
	\draw[thick, arrows=-stealth] (72pt,77pt) to[out=70,in=110] (89.5pt,76.5pt);
	
	\draw (92pt,0pt) -- (92pt,72pt);

	\draw[fill=black] (92pt,0pt) circle (0.9ex);
	\draw[fill=black] (92pt,18pt) circle (0.9ex);
	\draw[fill=black] (92pt,36pt) circle (0.9ex);
	\draw[fill=black] (92pt,54pt) circle (0.9ex);
	\draw[fill=black] (92pt,72pt) circle (0.9ex);
	
	\draw[densely dotted, thick, dgray] (85pt,78.5pt) -- (85pt,47.5pt) -- (99pt,47.5pt);
	\draw[densely dotted, thick, dgray] (99pt,47.5pt) -- (99pt,78.5pt) -- (85pt,78.5pt);

\end{tikzpicture}
			\vspace{-14pt}
		\captionsetup{justification=centering}
		\caption{}
		\label{fig:store_charge}
	\end{subfigure}
	\hspace{0.08\textwidth}
		\begin{subfigure}[b]{0.43\textwidth}
			\centering
			\definecolor{dgray}{gray}{0.4}

\begin{tikzpicture}

	\draw (0pt,0pt) -- (0pt,72pt);

	\draw[fill=black] (0pt,0pt) circle (0.9ex);
	\draw[fill=black] (0pt,18pt) circle (0.9ex);
	\draw[fill=black] (0pt,36pt) circle (0.9ex);
	\draw[fill=black] (0pt,54pt) circle (0.9ex);
	\draw[fill=black] (0pt,72pt) circle (0.9ex);
	
	\draw[densely dotted, thick, dgray] (-7pt,24.5pt) -- (-7pt,-6.5pt) -- (7pt,-6.5pt);
	\draw[densely dotted, thick, dgray] (7pt,-6.5pt) -- (7pt,24.5pt) -- (-7pt,24.5pt);
	
	\draw[thick, arrows=-stealth] (8pt,36pt) -- (16pt,36pt);
	
	\draw (23pt,0pt) -- (23pt,72pt);

	\draw[black, fill=white] (23pt,0pt) circle (0.9ex);
	\draw[fill=black] (23pt,18pt) circle (0.9ex);
	\draw[fill=black] (23pt,36pt) circle (0.9ex);
	\draw[fill=black] (23pt,54pt) circle (0.9ex);
	\draw[fill=black] (23pt,72pt) circle (0.9ex);
	
	\draw[densely dotted, thick, dgray] (16pt,42.5pt) -- (16pt,11.5pt) -- (30pt,11.5pt);
	\draw[densely dotted, thick, dgray] (30pt,11.5pt) -- (30pt,42.5pt) -- (16pt,42.5pt);
	
	\draw[thick, arrows=-stealth] (31pt,36pt) -- (39pt,36pt);
	
	\draw (46pt,0pt) -- (46pt,72pt);

	\draw[fill=black] (46pt,0pt) circle (0.9ex);
	\draw[black, fill=white] (46pt,18pt) circle (0.9ex);
	\draw[fill=black] (46pt,36pt) circle (0.9ex);
	\draw[fill=black] (46pt,54pt) circle (0.9ex);
	\draw[fill=black] (46pt,72pt) circle (0.9ex);
	
	\draw[densely dotted, thick, dgray] (39pt,60.5pt) -- (39pt,29.5pt) -- (53pt,29.5pt);
	\draw[densely dotted, thick, dgray] (53pt,29.5pt) -- (53pt,60.5pt) -- (39pt,60.5pt);
	
	\draw[thick, arrows=-stealth] (54pt,36pt) -- (62pt,36pt);
	
	\draw (69pt,0pt) -- (69pt,72pt);

	\draw[black, fill=white] (69pt,0pt) circle (0.9ex);
	\draw[fill=black] (69pt,18pt) circle (0.9ex);
	\draw[black, fill=white] (69pt,36pt) circle (0.9ex);
	\draw[fill=black] (69pt,54pt) circle (0.9ex);
	\draw[fill=black] (69pt,72pt) circle (0.9ex);
	
	\draw[densely dotted, thick, dgray] (62pt,78.5pt) -- (62pt,47.5pt) -- (76pt,47.5pt);
	\draw[densely dotted, thick, dgray] (76pt,47.5pt) -- (76pt,78.5pt) -- (62pt,78.5pt);
	
	\draw[thick, arrows=-stealth] (77pt,36pt) -- (85pt,36pt);
	
	\draw (92pt,0pt) -- (92pt,72pt);

	\draw[fill=black] (92pt,0pt) circle (0.9ex);
	\draw[black, fill=white] (92pt,18pt) circle (0.9ex);
	\draw[fill=black] (92pt,36pt) circle (0.9ex);
	\draw[black, fill=white] (92pt,54pt) circle (0.9ex);
	\draw[fill=black] (92pt,72pt) circle (0.9ex);

\end{tikzpicture}
			\vspace{-14pt}
		\captionsetup{justification=centering}
		\caption{}
		\label{fig:store_release}
	\end{subfigure}
	\caption{When charging (a), we propagate each new conflict to the next position (Figure \ref{fig:prop} shows the first step of (a) in detail). When unloading (b), we always propagate the lowermost stored conflict to the bottom.}\label{fig:allStore}
\hspace{0.0\textwidth}
\endminipage\hfill
\end{figure}

To provide a way to switch the control nodes in any order of our choice, we connect the levels of the construction with tools known as \textit{storage chains}. A storage chain is a path of 5 nodes, initially colored in an alternating fashion. The weights of the nodes in the chain are chosen such that each node is a follower node of its upper neighbor (this can be ensured by defining node weights in a bottom-to-top fashion, always choosing sufficiently large weight for the next node). The uppermost and lowermost nodes may have other upper and lower neighbors outside of the chain, respectively.

Assume now that the topmost node in the chain is switched by some external condition (i.e., its upper neighbors outside of the chain). This introduces a conflict into the chain between the uppermost two nodes, as shown in Figure \ref{fig:prop}. However, recall that by our definition of node weights, the second node (from the top) is a follower of the uppermost node, and therefore this conflict makes the second node switchable. Switching the second node (to black) resolves the original conflict, but creates a new conflict between the second and third nodes instead (now making the third node switchable). Generally, whenever there is a conflicting pair of subsequent nodes above an alternating-colored (part of the) chain, we are able to switch the lower node, and thus move the conflict down to the next node pair in the chain. We can use this method to move a conflict down to any point in the chain, as shown in the figure; we refer to this process as \textit{propagating down} the conflict in the chain.

With this technique, we can accumulate and store conflicts in the chain `for later use'. If the uppermost node is forced to switch 4 times, then we can propagate down each of the emerging conflicts to a different position (i.e., pair of nodes) in the chain, ending up with 4 conflicts in a completely monochromatic chain. This process (see Figure \ref{fig:store_charge}) is referred to as \textit{charging} the chain. In another sequence of steps, we can then \textit{unload} the chain and propagate these conflicts one by one to the bottom of the chain, essentially using the stored conflicts to switch the lowermost node 4 times in a timing of our own choice (see Figure \ref{fig:store_release}). When the sequence is finished, each node in the chain once again has its original color.

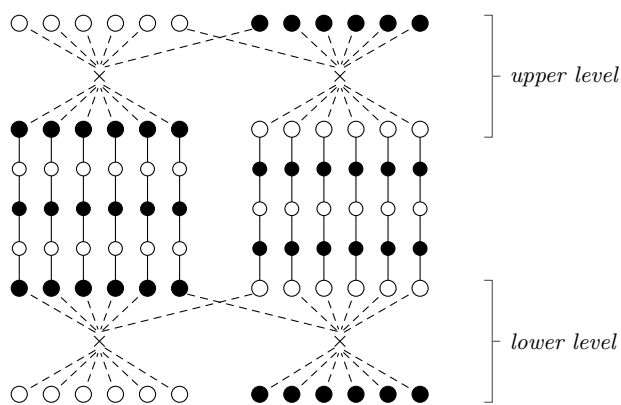
\begin{figure}
\hspace{0.26\textwidth}
\minipage{0.67\textwidth}
	\definecolor{dgray}{gray}{0.3}

\begin{tikzpicture}

	\draw (28pt,78pt) -- (32pt,82pt);
	\draw (28pt,82pt) -- (32pt,78pt);

	\draw[densely dashed] (0pt,60pt) -- (30pt,77.5pt);
	\draw[densely dashed] (12pt,60pt) -- (30pt,77pt);
	\draw[densely dashed] (24pt,60pt) -- (29.5pt,75.5pt);
	\draw[densely dashed] (36pt,60pt) -- (30.5pt,75.5pt);
	\draw[densely dashed] (48pt,60pt) -- (30pt,77pt);
	\draw[densely dashed] (60pt,60pt) -- (30pt,77.5pt);
	
	\draw[densely dashed] (0pt,100pt) -- (30pt,82.5pt);
	\draw[densely dashed] (12pt,100pt) -- (30pt,83pt);
	\draw[densely dashed] (24pt,100pt) -- (29.5pt,84.5pt);
	\draw[densely dashed] (36pt,100pt) -- (30.5pt,84.5pt);
	\draw[densely dashed] (48pt,100pt) -- (30pt,83pt);
	\draw[densely dashed] (92pt,99pt) -- (32pt,83.5pt);
	
	\draw (118pt,78pt) -- (122pt,82pt);
	\draw (118pt,82pt) -- (122pt,78pt);

	\draw[densely dashed] (90pt,60pt) -- (120pt,77.5pt);
	\draw[densely dashed] (102pt,60pt) -- (120pt,77pt);
	\draw[densely dashed] (114pt,60pt) -- (119.5pt,75.5pt);
	\draw[densely dashed] (126pt,60pt) -- (120.5pt,75.5pt);
	\draw[densely dashed] (138pt,60pt) -- (120pt,77pt);
	\draw[densely dashed] (150pt,60pt) -- (120pt,77.5pt);
	
	\draw[densely dashed] (58pt,99pt) -- (118pt,83.5pt);
	\draw[densely dashed] (102pt,100pt) -- (120pt,83pt);
	\draw[densely dashed] (114pt,100pt) -- (119.5pt,84.5pt);
	\draw[densely dashed] (126pt,100pt) -- (120.5pt,84.5pt);
	\draw[densely dashed] (138pt,100pt) -- (120pt,83pt);
	\draw[densely dashed] (150pt,100pt) -- (120pt,82.5pt);
	
	
	\draw (28pt,-18pt) -- (32pt,-22pt);
	\draw (28pt,-22pt) -- (32pt,-18pt);

	\draw[densely dashed] (0pt,0pt) -- (30pt,-17.5pt);
	\draw[densely dashed] (12pt,0pt) -- (30pt,-17pt);
	\draw[densely dashed] (24pt,0pt) -- (29.5pt,-15.5pt);
	\draw[densely dashed] (36pt,0pt) -- (30.5pt,-15.5pt);
	\draw[densely dashed] (48pt,0pt) -- (30pt,-17pt);
	\draw[densely dashed] (92pt,-1pt) -- (30pt,-16.5pt);
	
	\draw[densely dashed] (0pt,-40pt) -- (30pt,-22.5pt);
	\draw[densely dashed] (12pt,-40pt) -- (30pt,-23pt);
	\draw[densely dashed] (24pt,-40pt) -- (29.5pt,-24.5pt);
	\draw[densely dashed] (36pt,-40pt) -- (30.5pt,-24.5pt);
	\draw[densely dashed] (48pt,-40pt) -- (30pt,-23pt);
	\draw[densely dashed] (60pt,-40pt) -- (32pt,-22.5pt);
	
	\draw (118pt,-18pt) -- (122pt,-22pt);
	\draw (118pt,-22pt) -- (122pt,-18pt);

	\draw[densely dashed] (58pt,-1pt) -- (120pt,-16.5pt);
	\draw[densely dashed] (102pt,0pt) -- (120pt,-17pt);
	\draw[densely dashed] (114pt,0pt) -- (119.5pt,-15.5pt);
	\draw[densely dashed] (126pt,0pt) -- (120.5pt,-15.5pt);
	\draw[densely dashed] (138pt,0pt) -- (120pt,-17pt);
	\draw[densely dashed] (150pt,0pt) -- (120pt,-17.5pt);
	
	\draw[densely dashed] (90pt,-40pt) -- (118pt,-22.5pt);
	\draw[densely dashed] (102pt,-40pt) -- (120pt,-23pt);
	\draw[densely dashed] (114pt,-40pt) -- (119.5pt,-24.5pt);
	\draw[densely dashed] (126pt,-40pt) -- (120.5pt,-24.5pt);
	\draw[densely dashed] (138pt,-40pt) -- (120pt,-23pt);
	\draw[densely dashed] (150pt,-40pt) -- (120pt,-22.5pt);
	
	
	\draw (0pt,0pt) -- (0pt,60pt);

	\draw[fill=black] (0pt,0pt) circle (0.7ex);
	\draw[black, fill=white] (0pt,15pt) circle (0.6ex);
	\draw[fill=black] (0pt,30pt) circle (0.6ex);
	\draw[black, fill=white] (0pt,45pt) circle (0.6ex);
	\draw[fill=black] (0pt,60pt) circle (0.7ex);
	
	\draw (12pt,0pt) -- (12pt,60pt);

	\draw[fill=black] (12pt,0pt) circle (0.7ex);
	\draw[black, fill=white] (12pt,15pt) circle (0.6ex);
	\draw[fill=black] (12pt,30pt) circle (0.6ex);
	\draw[black, fill=white] (12pt,45pt) circle (0.6ex);
	\draw[fill=black] (12pt,60pt) circle (0.7ex);
	
	\draw (24pt,0pt) -- (24pt,60pt);

	\draw[fill=black] (24pt,0pt) circle (0.7ex);
	\draw[black, fill=white] (24pt,15pt) circle (0.6ex);
	\draw[fill=black] (24pt,30pt) circle (0.6ex);
	\draw[black, fill=white] (24pt,45pt) circle (0.6ex);
	\draw[fill=black] (24pt,60pt) circle (0.7ex);
	
	\draw (36pt,0pt) -- (36pt,60pt);

	\draw[fill=black] (36pt,0pt) circle (0.7ex);
	\draw[black, fill=white] (36pt,15pt) circle (0.6ex);
	\draw[fill=black] (36pt,30pt) circle (0.6ex);
	\draw[black, fill=white] (36pt,45pt) circle (0.6ex);
	\draw[fill=black] (36pt,60pt) circle (0.7ex);
	
	\draw (48pt,0pt) -- (48pt,60pt);

	\draw[fill=black] (48pt,0pt) circle (0.7ex);
	\draw[black, fill=white] (48pt,15pt) circle (0.6ex);
	\draw[fill=black] (48pt,30pt) circle (0.6ex);
	\draw[black, fill=white] (48pt,45pt) circle (0.6ex);
	\draw[fill=black] (48pt,60pt) circle (0.7ex);
	
	\draw (60pt,0pt) -- (60pt,60pt);

	\draw[fill=black] (60pt,0pt) circle (0.7ex);
	\draw[black, fill=white] (60pt,15pt) circle (0.6ex);
	\draw[fill=black] (60pt,30pt) circle (0.6ex);
	\draw[black, fill=white] (60pt,45pt) circle (0.6ex);
	\draw[fill=black] (60pt,60pt) circle (0.7ex);
	
	
	\draw (90pt,0pt) -- (90pt,60pt);

	\draw[black, fill=white] (90pt,0pt) circle (0.7ex);
	\draw[fill=black] (90pt,15pt) circle (0.6ex);
	\draw[black, fill=white] (90pt,30pt) circle (0.6ex);
	\draw[fill=black] (90pt,45pt) circle (0.6ex);
	\draw[black, fill=white] (90pt,60pt) circle (0.7ex);
	
	\draw (102pt,0pt) -- (102pt,60pt);

	\draw[black, fill=white] (102pt,0pt) circle (0.7ex);
	\draw[fill=black] (102pt,15pt) circle (0.6ex);
	\draw[black, fill=white] (102pt,30pt) circle (0.6ex);
	\draw[fill=black] (102pt,45pt) circle (0.6ex);
	\draw[black, fill=white] (102pt,60pt) circle (0.7ex);
	
	\draw (114pt,0pt) -- (114pt,60pt);

	\draw[black, fill=white] (114pt,0pt) circle (0.7ex);
	\draw[fill=black] (114pt,15pt) circle (0.6ex);
	\draw[black, fill=white] (114pt,30pt) circle (0.6ex);
	\draw[fill=black] (114pt,45pt) circle (0.6ex);
	\draw[black, fill=white] (114pt,60pt) circle (0.7ex);
	
	\draw (126pt,0pt) -- (126pt,60pt);

	\draw[black, fill=white] (126pt,0pt) circle (0.7ex);
	\draw[fill=black] (126pt,15pt) circle (0.6ex);
	\draw[black, fill=white] (126pt,30pt) circle (0.6ex);
	\draw[fill=black] (126pt,45pt) circle (0.6ex);
	\draw[black, fill=white] (126pt,60pt) circle (0.7ex);
	
	\draw (138pt,0pt) -- (138pt,60pt);

	\draw[black, fill=white] (138pt,0pt) circle (0.7ex);
	\draw[fill=black] (138pt,15pt) circle (0.6ex);
	\draw[black, fill=white] (138pt,30pt) circle (0.6ex);
	\draw[fill=black] (138pt,45pt) circle (0.6ex);
	\draw[black, fill=white] (138pt,60pt) circle (0.7ex);
	
	\draw (150pt,0pt) -- (150pt,60pt);

	\draw[black, fill=white] (150pt,0pt) circle (0.7ex);
	\draw[fill=black] (150pt,15pt) circle (0.6ex);
	\draw[black, fill=white] (150pt,30pt) circle (0.6ex);
	\draw[fill=black] (150pt,45pt) circle (0.6ex);
	\draw[black, fill=white] (150pt,60pt) circle (0.7ex);
	
	
	\draw[black, fill=white] (0pt,100pt) circle (0.7ex);
	\draw[black, fill=white] (12pt,100pt) circle (0.7ex);
	\draw[black, fill=white] (24pt,100pt) circle (0.7ex);
	\draw[black, fill=white] (36pt,100pt) circle (0.7ex);
	\draw[black, fill=white] (48pt,100pt) circle (0.7ex);
	\draw[black, fill=white] (60pt,100pt) circle (0.7ex);
	
	\draw[fill=black] (90pt,100pt) circle (0.7ex);
	\draw[fill=black] (102pt,100pt) circle (0.7ex);
	\draw[fill=black] (114pt,100pt) circle (0.7ex);
	\draw[fill=black] (126pt,100pt) circle (0.7ex);
	\draw[fill=black] (138pt,100pt) circle (0.7ex);
	\draw[fill=black] (150pt,100pt) circle (0.7ex);
	
	\draw[black, fill=white] (0pt,-40pt) circle (0.7ex);
	\draw[black, fill=white] (12pt,-40pt) circle (0.7ex);
	\draw[black, fill=white] (24pt,-40pt) circle (0.7ex);
	\draw[black, fill=white] (36pt,-40pt) circle (0.7ex);
	\draw[black, fill=white] (48pt,-40pt) circle (0.7ex);
	\draw[black, fill=white] (60pt,-40pt) circle (0.7ex);
	
	\draw[fill=black] (90pt,-40pt) circle (0.7ex);
	\draw[fill=black] (102pt,-40pt) circle (0.7ex);
	\draw[fill=black] (114pt,-40pt) circle (0.7ex);
	\draw[fill=black] (126pt,-40pt) circle (0.7ex);
	\draw[fill=black] (138pt,-40pt) circle (0.7ex);
	\draw[fill=black] (150pt,-40pt) circle (0.7ex);
	
	\draw[dgray] (174pt,103pt) -- (177pt,103pt) -- (177pt,57pt) -- (174pt,57pt);
	\draw[dgray] (177pt,80pt) -- (180pt,80pt);
	\node[anchor=west] at (180pt,80pt) {\footnotesize \textit{upper level}};
	
	\draw[dgray] (174pt,3pt) -- (177pt,3pt) -- (177pt,-43pt) -- (174pt,-43pt);
	\draw[dgray] (177pt,-20pt) -- (180pt,-20pt);
	\node[anchor=west] at (180pt,-20pt) {\footnotesize \textit{lower level}};

\end{tikzpicture}
\endminipage\hfill
	\caption{Two levels of the construction, connected by storage chains (edges within a level are shown in dashed). For simpler illustration, the two sides of the lower level are horizontally swapped.}
	\label{fig:level}
\end{figure}

We use such storage chains to connect subsequent levels of our construction, with the base nodes and control nodes being the uppermost and lowermost nodes in the chains, respectively, as shown in Figure \ref{fig:level}.
This way, every time after the 6-tuple of base nodes in the upper level switch (together), we can execute the next step in charging each of the storage chains. After each of the base nodes switch 4 times, each of the storage chains are charged. Then, by unloading each chain in 4 steps in the order of our choice, we can switch each of the control nodes below 4 times, in any preferred order; this enables us to execute the control sequence on the lower level. Thus, if the upper-level base nodes are switched 4 times, we can indeed switch the lower-level base nodes 6 times.

For a high-level overview of the process, the execution of the adversarial sequence on a given level $L$ could be summarized by the following recursive pseudocode:

\vspace{-5pt}

\begin{algorithm}[H]
{\color{darkgray} \hrule height 0.8pt}
\vspace{4pt}
\begin{algorithmic}
\Function{ProcessLevel}{$L$}
\For{each of the 6 steps of the control sequence:}
\State On both sides, switch the next subset of 4 control nodes
\State Switch all $6+6$ base nodes
\State Propagate down the conflict in each chain as far as possible
\If{the chains below are fully charged:} 
\State Call \textsc{ProcessLevel}($L+1$) \footnotesize (\textsl{execution continues on level below}) \normalsize
\EndIf
\EndFor
\vspace{-3pt}
\State \Return \footnotesize (\textsl{execution continues on level above}) \normalsize
\EndFunction

\vspace{4pt}
{\color{darkgray} \hrule height 0.8pt}
\end{algorithmic}
\end{algorithm}

\vspace{-20pt}

Even with the storage chain connections, the addition of each new level increases the number of nodes only by a constant value. This implies that a graph on $n$ nodes can contain $\Theta(n)$ levels, and thus each node in the lowermost level indeed switches $2^{\Theta(n)}$ times.

There is one more detail to discuss: for convenience, we assumed that the number of switches $s$ in an upper level is always divisible by 4. However, $s$ switches in each control node in fact allows for only $\lfloor \frac{s}{4} \rfloor$ complete executions of the control sequence, and hence $\lfloor \frac{s}{4} \rfloor \cdot 6$ switches for the base nodes. Nonetheless, this still implies exponential increase for $s$ large enough (for example, $\lfloor \frac{s}{4} \rfloor \cdot 6 \geq \frac{6}{5} s$ holds if $s \geq 20$). Thus to overcome this problem, we ensure that the control nodes in the uppermost level already switch 20 times; this is achieved by adding an initially charged storage chain of 21 nodes above each uppermost control node. Unloading the chains allows us to switch these top-level control nodes 20 times in the preferred order, and thus the exponential increase of switches is guaranteed.

This proves our lower bound in model $SA$ for the case of Rule II with $\Lambda=5-\epsilon$ for any $\epsilon>0$. However, the construction is straightforward to generalize to any other odd integer $\Lambda_B$: for most of the analysis, one only needs to replace the value 4 by ($\Lambda_B-1$) and the value 6 by ($\Lambda_B+1$). This provides a construction with $(\Lambda_B+1)$-tuples of control and base nodes, and a $\frac{\Lambda_B+1}{\Lambda_B-1}$ factor of increase in switches for every new level. The control sequence can also be generalized for other $\Lambda_B$ values; details of the generalization are discussed in Appendix \ref{App:C}.

\section{Benevolent Case} \label{sec:ben}

It is significantly more difficult to show an exponential lower bound for benevolent models, since such a construction needs to guarantee that every possible sequence lasts for an exponential number of steps. We overcome this problem by heavily restricting the set of selectable sequences in the graph. Specifically, we start from the construction of Section \ref{sec:adv}, and we show how to add a set of extra nodes which ensure that the previously defined sequence is the only possible sequence the benevolent player can choose. In this section, we outline the main ideas of this benevolent construction; Appendix \ref{App:A} provides a detailed discussion of the technique.

\begin{theorem}
For Switching Rule II with any $\lambda < 1$, there exists a class of (sparse) weighted graphs that have $2^{\Theta(n)}$ stabilization time in the benevolent models (models $SB$ and $CB$).
\end{theorem}

We basically use two tools (gadgets) to ensure that the player, when selecting the sequence, has to follow the procedure described in the pseudocode above. On the one hand, we show how to build logical \textit{\textsc{and} gates} and \textit{\textsc{or} gates}, in order to check that a given step of the procedure is reached, and use these gates to allow the player to proceed to the next step of the procedure. On the other hand, we devise a \textit{state chain} in order to keep track of the current phase of the procedure, which can then be used as a condition in the logical gates that control the execution of the procedure.

With the appropriate combination of these two gadgets, we can ensure that the benevolent player has no other option than to switch the control nodes, base nodes and storage chain nodes in the order described by the recursive procedure. We add a separate such combination of these gadgets to each level of the construction of Section \ref{sec:adv}. However, since in our recursive procedure, each level of the graph executes the same sequence of steps multiple times (the lower levels exponentially many times), the design of these gadgets also needs to ensure that the gadget can execute its task multiple times. This is achieved through introducing a method to repeatedly `reset' the gadgets to their initial state.

For the purpose of resetting these gadgets, we introduce another tool, the third main ingredient of our benevolent construction, known as a \textit{pacer system}. The main idea of the resetting technique is to connect each gadget (logical gate or state chain) to so-called \textit{pacer nodes} higher in the graph, and to ensure that each such pacer node switches at least twice between two consecutive times of using the gadget. The gadgets are designed in a way which guarantees that this pacer node switching twice results in the gadget being reset to its default state (i.e., each node to its initial color).

Such a pacer node essentially `recharges' the gadget with conflicts: since the weighted sum of conflicts in the graph monotonically decreases, the gadget can only return to the same (initial) state repeatedly if it `acquires' new conflicts from some other part of the graph. This is achieved through the connection to the pacer node, which is in a higher level of the graph (with larger weights), and thus has significantly more conflicts to `push down' into the gadget as a byproduct of its switching.

The simplest way to add pacer nodes to our construction is to place a pair of them between a set of control and base nodes, as shown in Figure \ref{fig:pacer_base}. In this modified level version, the steps of the control sequence do not switch the base nodes directly. Instead, this happens indirectly: after 5 of the 6 control nodes are black, first the upper pacer node, and then the lower pacer node switches, followed by the base nodes in the end. Thus, the addition of pacer nodes leaves the general behavior of the level unchanged: the base nodes will still switch eventually after each step of the control sequence. However, in this new level construction, the newly added pacer nodes will also both switch in each of these steps.

The actual pacer systems used in our construction, discussed in Appendix \ref{App:A}, are more sophisticated constructions based on this idea. They consist of multiple pacer nodes in order to be able to recharge gadgets of both colors, and they are also responsible for checking that the recharging process has indeed been executed on the connected gadgets.

Given the technique to reset gadgets, it only remains to briefly present the behavior of the two gadgets (logical gates and state chains), and to outline how they are used in the construction. For the convenient description of gadgets, we first introduce two special kinds of node concepts. Essentially, these are methods to carefully select the weight of some specific neighbors of nodes such that they fulfill the following roles:

\vspace{-6pt}

\begin{itemize}

  \item \textit{Observer node}: given a set of nodes $U_0$, we can add a new common neighbor $v_o$ to these nodes such that the behavior of $v_o$ depends on the nodes in $U_0$, but the behavior of $U_0$ is unaffected by the addition of $v_o$
	
\vspace{5pt}
	
  \item \textit{Enabler node}: given a node $u_1$ dominated by another node $u_d$, we can add a new neighbor $v_e$ to $u_1$, such that $u_1$ is no longer dominated by $\{ u_d \}$, but it is dominated by the subset $\{ u_d, v_e \}$
	
\end{itemize}

\vspace{-2pt}

These techniques and the properties of such nodes are discussed in detail in Appendix \ref{App:A}.

Given a set of input nodes $U_0$ and an output node $u_1$, we can use these concepts to build an \textsc{and} gate which only enables the switching of $u_1$ if all nodes in $U_0$ are colored with a given color. This gadget connects to each of the input nodes in $U_0$ through a common observer node, and connects to the output $u_1$ through an enabler node. Besides the observer and enabler node, the gadget only requires an extra relay node (and an appropriate choice of weights) to connect these two nodes, and an extra upper neighbor for each node in order to connect the gadget to a pacer system which resets it after use. A brief illustration of the gadget is available in Figure \ref{fig:and}, with the details discussed in Appendix \ref{App:A}. In a very similar fashion, we can also create \textsc{and} gates for inputs of the other color, \textsc{or} gates, or even multi-layer gates that allow us to combine different conditions.

\begin{figure}
\centering
\hspace{0.01\textwidth}
\minipage{0.44\textwidth}
	\centering
	\vspace{25pt}
	\begin{tikzpicture}

	\draw (0pt,18pt) -- (-35pt,0pt);
	\draw (0pt,18pt) -- (-21pt,0pt);
	\draw (0pt,18pt) -- (-7pt,0pt);
	\draw (0pt,18pt) -- (7pt,0pt);
	\draw (0pt,18pt) -- (21pt,0pt);
	\draw (0pt,18pt) -- (35pt,0pt);
	
	\draw (0pt,33pt) -- (-35pt,51pt);
	\draw (0pt,33pt) -- (-21pt,51pt);
	\draw (0pt,33pt) -- (-7pt,51pt);
	\draw (0pt,33pt) -- (7pt,51pt);
	\draw (0pt,33pt) -- (21pt,51pt);
	\draw (0pt,33pt) -- (35pt,51pt);
	
	\draw (0pt,18pt) -- (0pt,33pt);
	
	\draw[black, fill=white] (-35pt,51pt) circle (0.7ex);
	\draw[black, fill=white] (-21pt,51pt) circle (0.7ex);
	\draw[black, fill=white] (-7pt,51pt) circle (0.7ex);
	\draw[black, fill=white] (7pt,51pt) circle (0.7ex);
	\draw[fill=black] (21pt,51pt) circle (0.7ex);
	\draw[black, fill=white] (35pt,51pt) circle (0.7ex);
	
	\draw[black, fill=white] (0pt,18pt) circle (0.9ex);
	\draw[fill=black] (0pt,33pt) circle (0.9ex);
	
	\draw[fill=black] (-35pt,0pt) circle (0.7ex);
	\draw[fill=black] (-21pt,0pt) circle (0.7ex);
	\draw[fill=black] (-7pt,0pt) circle (0.7ex);
	\draw[fill=black] (7pt,0pt) circle (0.7ex);
	\draw[fill=black] (21pt,0pt) circle (0.7ex);
	\draw[fill=black] (35pt,0pt) circle (0.7ex);
	
\end{tikzpicture}
	\caption{Adding a pair of pacers between a layer of control nodes and base nodes.}
	\label{fig:pacer_base}
\endminipage\hfill
\hspace{0.02\textwidth}
\minipage{0.42\textwidth}
	\centering
	\definecolor{dgray}{gray}{0.25}

\begin{tikzpicture}

	\draw (0pt,0pt) -- (50pt,0pt);
	\draw (0pt,0pt) -- (0pt,30pt);
	\draw (25pt,0pt) -- (25pt,30pt);
	\draw (50pt,0pt) -- (50pt,30pt);
	
	\draw[dashed] (0pt,0pt) -- (-30pt,15pt);
	\draw[dashed] (0pt,0pt) -- (-30pt,0pt);
	\draw[dashed] (0pt,0pt) -- (-30pt,-15pt);
	\draw[dashed] (50pt,0pt) -- (80pt,0pt);

	\draw[fill=black] (0pt,0pt) circle (0.9ex);	
	\draw[black, fill=white] (25pt,0pt) circle (0.9ex);
	\draw[fill=black] (50pt,0pt) circle (0.9ex);
	
	\draw[fill=black] (0pt,30pt) circle (0.9ex);	
	\draw[black, fill=white] (25pt,30pt) circle (0.9ex);
	\draw[fill=black] (50pt,30pt) circle (0.9ex);
	
	\draw[black, fill=white] (-30pt,15pt) circle (0.7ex);	
	\draw[fill=black] (-30pt,0pt) circle (0.7ex);
	\draw[black, fill=white] (-30pt,-15pt) circle (0.7ex);
	
	\draw[gray] (-35pt,20pt) -- (-37pt,20pt) -- (-37pt,-20pt) -- (-35pt,-20pt);
	\draw[gray] (-37pt,0pt) -- (-39pt,0pt);
	\node[anchor=south, rotate=90] at (-38pt,0pt) {\scriptsize \textit{nodes}};
	\node[anchor=south, rotate=90] at (-45pt,0pt) {\scriptsize \textit{observed}};
	
	\draw[thick, gray, densely dotted] (-7pt,23pt) -- (-7pt,37pt) -- (58pt,37pt) -- (58pt,23pt) -- cycle;
	\draw[gray] (25pt,37pt) -- (25pt,40pt);
	\node[anchor=south] at (25pt,37pt) {\scriptsize \textit{pacers}};
	
	\draw[black, fill=white] (80pt,0pt) circle (0.7ex);
	\node[anchor=north, rotate=90] at (82pt,0pt) {\scriptsize \textit{enabled}};
	\node[anchor=north, rotate=90] at (89pt,0pt) {\scriptsize \textit{node}};
	
	\node[anchor=north] at (1pt,-6pt) {\scriptsize \textit{observer}};
	\node[anchor=north] at (50pt,-6pt) {\scriptsize \textit{enabler}};
	
\end{tikzpicture}
	\caption{Logical (e.g. \textsc{and}) gate.}
	\label{fig:and}
\endminipage\hfill
\hspace{0.9999\textwidth}
\end{figure}

\begin{figure}
\hspace{0.26\textwidth}
\minipage{0.67\textwidth}
	\definecolor{dgray}{gray}{0.4}

\begin{tikzpicture}

	\draw (0pt,0pt) -- (0pt,72pt);

	\draw[fill=black] (0pt,0pt) circle (0.9ex);
	\draw[black, fill=white] (0pt,18pt) circle (0.9ex);
	\draw[fill=black] (0pt,36pt) circle (0.9ex);
	\draw[black, fill=white] (0pt,54pt) circle (0.9ex);
	\draw[black, fill=white] (0pt,72pt) circle (0.9ex);
	
	\draw[densely dotted, thick, dgray] (5pt,77.5pt) -- (9pt,77.5pt) -- (9pt,48.5pt) -- (5pt,48.5pt);
	\node[anchor=north] at (21pt,73pt) {\scriptsize \textit{first}};
	\node[anchor=north] at (21pt,66pt) {\scriptsize \textit{state}};
	
	\draw[thick, arrows=-stealth] (27pt,36pt) -- (43pt,36pt);
	
	\draw (60pt,0pt) -- (60pt,72pt);

	\draw[fill=black] (60pt,0pt) circle (0.9ex);
	\draw[black, fill=white] (60pt,18pt) circle (0.9ex);
	\draw[fill=black] (60pt,36pt) circle (0.9ex);
	\draw[fill=black] (60pt,54pt) circle (0.9ex);
	\draw[black, fill=white] (60pt,72pt) circle (0.9ex);
	
	\draw[densely dotted, thick, dgray] (65pt,59.5pt) -- (69pt,59.5pt) -- (69pt,30.5pt) -- (65pt,30.5pt);
	\node[anchor=north] at (84pt,55pt) {\scriptsize \textit{second}};
	\node[anchor=north] at (84pt,48pt) {\scriptsize \textit{state}};
	
	\draw[thick, arrows=-stealth] (87pt,36pt) -- (103pt,36pt);
	
	\draw (120pt,0pt) -- (120pt,72pt);

	\draw[fill=black] (120pt,0pt) circle (0.9ex);
	\draw[black, fill=white] (120pt,18pt) circle (0.9ex);
	\draw[black, fill=white] (120pt,36pt) circle (0.9ex);
	\draw[fill=black] (120pt,54pt) circle (0.9ex);
	\draw[black, fill=white] (120pt,72pt) circle (0.9ex);
	
	\draw[densely dotted, thick, dgray] (125pt,41.5pt) -- (129pt,41.5pt) -- (129pt,12.5pt) -- (125pt,12.5pt);
	\node[anchor=north] at (141pt,37pt) {\scriptsize \textit{third}};
	\node[anchor=north] at (141pt,30pt) {\scriptsize \textit{state}};
	
	\draw[thick, arrows=-stealth] (147pt,36pt) -- (163pt,36pt);
	
	\draw (180pt,0pt) -- (180pt,72pt);

	\draw[fill=black] (180pt,0pt) circle (0.9ex);
	\draw[fill=black] (180pt,18pt) circle (0.9ex);
	\draw[black, fill=white] (180pt,36pt) circle (0.9ex);
	\draw[fill=black] (180pt,54pt) circle (0.9ex);
	\draw[black, fill=white] (180pt,72pt) circle (0.9ex);
	
	\draw[densely dotted, thick, dgray] (185pt,23.5pt) -- (189pt,23.5pt) -- (189pt,-5.5pt) -- (185pt,-5.5pt);
	\node[anchor=north] at (202pt,19pt) {\scriptsize \textit{fourth}};
	\node[anchor=north] at (202pt,12pt) {\scriptsize \textit{state}};

\end{tikzpicture}
\endminipage\hfill
	\caption{Simplified illustration of a state chain on 4 states (see Appendix \ref{App:A} for a more detailed illustration). The position of the conflict in the chain shows our current state of the procedure. When propagating the conflict down by one step, the chain proceeds to the next state.}
	\label{fig:chainstates}
\end{figure}
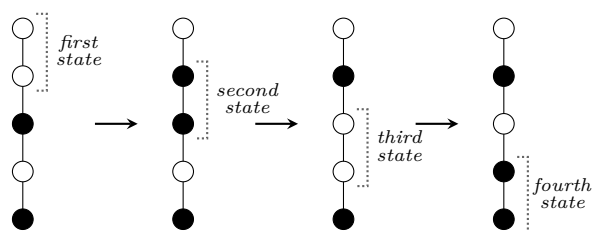

Besides logical gates, the other key gadget in our benevolent construction is the state chain. For each level of the construction, we add a separate state chain in order to indicate the current state (i.e., point in the execution) of the procedure on this level. Essentially, a state chain is a vertical chain of nodes, where every node in the chain is dominated by its upper neighbor, similarly to the case of a storage chain. However, while storage chains are used to accumulate conflicts, a state chain will, on the other hand, always contain exactly one conflict, which we propagate down step by step. The different possible positions of the conflict can then correspond to different states of the procedure, and at a given point in time, our current state in the procedure is determined by the current position of the conflict in the chain (as illustrated in Figure \ref{fig:chainstates}).

One such state chain is added to each level of our benevolent construction. The node pairs in the chain that express a state are included in the conditions of the logical gates that control the execution of the recursive procedure on the level, ensuring that certain steps are only available to the benevolent player at certain points in the process. Furthermore, the nodes in the state chain are also connected to enabler nodes, and thus proceeding to the next state is always based on a given condition. Therefore, the benevolent player has no other option than to simultaneously proceed through the steps of the recursive process and the states of the state chain, in the appropriate order. With a couple of auxiliary nodes at the top of the chain, we can also connect the state chain to a pacer node, allowing us to reset the chain and jump back to the first state whenever the last state of the chain is reached.

Given these gadgets, let us now briefly reflect on the states and conditions we need to encode in order to ensure that the player has to follow the recursive sequence. The main idea is to use the logical gates to control the flow of execution within a given level: through the enabler nodes of the gates, we ensure that the switching of the next $2 \times 4$ control nodes (i.e., the next step of the control sequence) is only enabled after the previous switching of the base nodes is finished. In practice, this means that, after the base nodes have switched, when the newly added conflicts are propagated down far enough in each of the $2 \times 6$ storage chains below, the gates enable the further down-propagation of the appropriate $2 \times 4$ conflicts in the storage chains above, which will in turn make the next subset of $2 \times 4$ control nodes switchable. That is, the input (observer) nodes of these logical gates are connected to specific nodes of the storage chains below the level, while their output (enabler) nodes are connected to nodes of the storage chains above the level.

However, recall that charging the storage chains below takes 4 steps, while executing the control sequence above consist of 6 steps, so the two processes do not remain in synchrony. Thus in different phases of the procedure, the same set of storage chain nodes below have to enable different subsets of the control nodes above. Because of this, our construction encodes these different phases of the procedure as states in the state chain, and the appropriate state is also included in the condition of the logical gate that enables the next set of control nodes. When a cycle is finished (i.e., the two processes return to their default state at the same time), the state chain is reset and iteration starts again from the first state of the chain.

Furthermore, note that throughout the recursion, execution repeatedly leaves the current level and continues on the level above (or below), so the state chain of each level also has specific states indicating that the execution is currently on a level above (or below).

Altogether, these benevolent-case modifications only add constantly many gadgets (each of constant-size) to each level of the construction. Therefore, the modified construction still has only $O(1)$ nodes in a level, allowing for $\Theta(n)$ levels and thus $2^{\Theta(n)}$ stabilization time. This establishes our lower bound for model $SB$. By design, the construction only has a few (at most constantly many) switchable nodes at every point in time, and thus even in model $CB$, it allows for only very limited concurrency for the benevolent player. Specifically, since there are concrete nodes in the construction that switch $2^{\Theta(n)}$ times, the number of steps is still exponential in model $CB$.

Also, note that even with the gadgets added in the benevolent case, each node of the graph still has a constant degree, and thus our bound is also valid for sparse graphs.

\bibliography{references}

\begin{thebibliography}{10}

\bibitem{aharoniUGP}
Ron Aharoni, Eric~C Milner, and Karel Prikry.
\newblock Unfriendly partitions of a graph.
\newblock {\em Journal of Combinatorial Theory, Series B}, 50(1):1--10, 1990.

\bibitem{SGPclass5}
Cristina Bazgan, Zsolt Tuza, and Daniel Vanderpooten.
\newblock On the existence and determination of satisfactory partitions in a
  graph.
\newblock In {\em International Symposium on Algorithms and Computation}, pages
  444--453. Springer, 2003.

\bibitem{approx0}
Cristina Bazgan, Zsolt Tuza, and Daniel Vanderpooten.
\newblock Complexity and approximation of satisfactory partition problems.
\newblock In {\em International Computing and Combinatorics Conference}, pages
  829--838. Springer, 2005.

\bibitem{SGPclass4}
Cristina Bazgan, Zsolt Tuza, and Daniel Vanderpooten.
\newblock The satisfactory partition problem.
\newblock {\em Discrete applied mathematics}, 154(8):1236--1245, 2006.

\bibitem{SGPsurvey}
Cristina Bazgan, Zsolt Tuza, and Daniel Vanderpooten.
\newblock Satisfactory graph partition, variants, and generalizations.
\newblock {\em European Journal of Operational Research}, 206(2):271--280,
  2010.

\bibitem{applic1}
Olivier Bodini, Thomas Fernique, and Damien Regnault.
\newblock Quasicrystallization by stochastic flips.
\newblock {\em HAL online archives}, 2009.

\bibitem{applic2}
Olivier Bodini, Thomas Fernique, and Damien Regnault.
\newblock Stochastic flips on two-letter words.
\newblock In {\em 2010 Proceedings of the Seventh Workshop on Analytic
  Algorithmics and Combinatorics (ANALCO)}, pages 48--55. SIAM, 2010.

\bibitem{UGPrayless}
Henning Bruhn, Reinhard Diestel, Agelos Georgakopoulos, and Philipp
  Spr{\"u}ssel.
\newblock Every rayless graph has an unfriendly partition.
\newblock {\em Combinatorica}, 30(5):521--532, 2010.

\bibitem{applic4}
Zhigang Cao and Xiaoguang Yang.
\newblock The fashion game: Network extension of matching pennies.
\newblock {\em Theoretical Computer Science}, 540:169--181, 2014.

\bibitem{applic3}
Jacques Demongeot, Julio Aracena, Florence Thuderoz, Thierry-Pascal Baum, and
  Olivier Cohen.
\newblock Genetic regulation networks: circuits, regulons and attractors.
\newblock {\em Comptes Rendus Biologies}, 326(2):171--188, 2003.

\bibitem{majority}
Silvio Frischknecht, Barbara Keller, and Roger Wattenhofer.
\newblock Convergence in (social) influence networks.
\newblock In {\em International Symposium on Distributed Computing}, pages
  433--446. Springer, 2013.

\bibitem{SGPclass1}
Michael~U Gerber and Daniel Kobler.
\newblock Algorithmic approach to the satisfactory graph partitioning problem.
\newblock {\em European Journal of Operational Research}, 125(2):283--291,
  2000.

\bibitem{SGPclass2}
Michael~U Gerber and Daniel Kobler.
\newblock Classes of graphs that can be partitioned to satisfy all their
  vertices.
\newblock {\em Australasian Journal of Combinatorics}, 29:201--214, 2004.

\bibitem{hedetniemi}
Sandra~M Hedetniemi, Stephen~T Hedetniemi, KE~Kennedy, and Alice~A Mcrae.
\newblock Self-stabilizing algorithms for unfriendly partitions into two
  disjoint dominating sets.
\newblock {\em Parallel Processing Letters}, 23(01):1350001, 2013.

\bibitem{majorityW}
Barbara Keller, David Peleg, and Roger Wattenhofer.
\newblock How even tiny influence can have a big impact!
\newblock In {\em International Conference on Fun with Algorithms}, pages
  252--263. Springer, 2014.

\bibitem{KPRanticoor}
Jeremy Kun, Brian Powers, and Lev Reyzin.
\newblock Anti-coordination games and stable graph colorings.
\newblock In {\em International Symposium on Algorithmic Game Theory}, pages
  122--133. Springer, 2013.

\bibitem{CA1}
Damien Regnault, Nicolas Schabanel, and {\'E}ric Thierry.
\newblock Progresses in the analysis of stochastic 2d cellular automata: A
  study of asynchronous 2d minority.
\newblock In Lud{\v{e}}k Ku{\v{c}}era and Anton{\'i}n Ku{\v{c}}era, editors,
  {\em Mathematical Foundations of Computer Science 2007}, pages 320--332.
  Springer Berlin Heidelberg, 2007.

\bibitem{CA2}
Damien Regnault, Nicolas Schabanel, and {\'E}ric Thierry.
\newblock On the analysis of “simple” 2d stochastic cellular automata.
\newblock In {\em International Conference on Language and Automata Theory and
  Applications}, pages 452--463. Springer, 2008.

\bibitem{CA3}
Jean-Baptiste Rouquier, Damien Regnault, and {\'E}ric Thierry.
\newblock Stochastic minority on graphs.
\newblock {\em Theoretical Computer Science}, 412(30):3947--3963, 2011.

\bibitem{SGPclass3}
Khurram~H Shafique and Ronald~D Dutton.
\newblock On satisfactory partitioning of graphs.
\newblock {\em Congressus Numerantium}, pages 183--194, 2002.

\bibitem{noUGP}
Saharon Shelah and Eric~C Milner.
\newblock Graphs with no unfriendly partitions.
\newblock {\em A tribute to Paul Erd{\"o}s}, pages 373--384, 1990.

\end{thebibliography}

\newpage

\begin{appendices}

\section{Details of the Benevolent Construction} \label{App:A}

We now describe the benevolent construction in more detail. Note that some of the previous presentation conventions from Section \ref{sec:adv} are not followed when presenting gadgets: for example, nodes shown at the same vertical position do not necessarily have the same weight. Also, given a node $v$ that is originally following a node $v_d$, we often refer to $v_d$ as the dominant node of $v$ even after adding an enabler neighbor to $v$, although technically $v$ is not dominated by $\{ v_d \}$ anymore.

While extending the construction of Section \ref{sec:adv} to the benevolent case, we often apply the step of increasing the weight of a dominant node to a sufficiently large value to ensure that the follower node behaves in a certain way. Note that this is a step that we can always execute: besides increasing the dominant node by a given factor, we also increase the weight of each direct and indirect upper neighbor of the node by the same factor. This guarantees that the behavior of each node above remains unchanged.

\paragraph*{Enabler nodes}

Assume we have a follower node $v$ with a dominant neighbor $v_d$. For convenience, let us introduce the notation $w_R:=W_{N(v) \setminus \{ v_d \} }$ for the weight of the remaining nodes. The goal of adding an enabler node $v_e$ is to ensure that $v_d$ alone does not dominate $v$ anymore, but $\{v_d, v_e\}$ dominates $v$. Since $v$ is a follower node, we initially have $w(v_d) \geq \Lambda \cdot w_R$. We need to select $w(v_e) > w_R$ to ensure that $\{v_d, v_e\}$ has larger weight than the original neighborhood. Since we know that $w_R < \frac{1}{\Lambda+1} W_{N(v)}$ holds, a choice of, say, $w(v_e) := \frac{1}{\Lambda} W_{N(v)}$ suffices.

We then have to assign a new weight $w'$ to $v_d$ such that $\{v_d, v_e\}$ dominates $v$, but the original set of nodes ($\{v_d\}$ with the nodes composing $w_R$) does not. These two conditions require $w'+w(v_e) \geq \Lambda \cdot w_R$ and $w'+ w_R < \Lambda \cdot w(v_e)$, respectively. This implies that a choice of $w' \in \left[ \Lambda w_R - w(v_e) ; \, \Lambda w(v_e) - w_R  \right)$ suffices; this is possible, since we have specifically chosen $w(v_e) >  w_R$.

An additional idea also allows us to create an \textit{asymmetric enabler} node $v_e$, such that $v_d$ alone being black is enough to force $v$ to turn white, but we need both $v_d$ and $v_e$ to be white in order to force $v$ to turn black (or vice versa). In such a situation, we say that $v_d$ semi-dominates $v$ for one color (black in this case). This can be achieved by choosing $w(v_e) = \frac{1}{\Lambda} W_{N(v)}$ as before, and also adding a fixed black neighbor to $v$ with the same weight $w(v_e)$. If we now set $w'$ such that $w'+w(v_e)$ is larger than $\frac{\Lambda}{\Lambda+1}$ times the sum of new weights of $v$'s neighbors, but $w'$ alone is not, then it requires $v_d$ and at least one of the other two large-weight nodes (i.e., $v_e$ or the fixed node) to force $v$ to switch. Since the fixed neighbor will always remain black, this implies that we have created an enabler node that is indeed only required in one of the directions for switching $v$.

\paragraph*{General enabler sets}

In a more general setting, we can add a set of $m$ enabler nodes $v_{e_1}$, ..., $v_{e_m}$ (to the same node $v$), such that for a given threshold $t \leq m$, $v$ becomes switchable only if both $v_d$ and at least $t$ out of the $m$ enabler nodes are of the same color. For this, we once again choose $w(v_{e_i}) = \frac{1}{\Lambda} W_{N(v)}$ for each of these enabler nodes to ensure that $w(v_{e_i}) > w_R$; for simplicity, let us write $w_E:=w(v_{e_i})$. Then for the desired behavior we have to ensure $w'+t w_E \geq \Lambda \cdot ((m-t) \cdot w_E + w_R)$ and $w'+(t-1) \cdot w_E + w_R < \Lambda \cdot (m-t+1) \cdot w_E$, which is possible with the appropriate choice of $w'$ as before.

However, we also have to ensure that the enabler nodes cannot dominate $v$ without the help of $v_d$; that is, we need $m \cdot w_E + w_R < \Lambda w'$, which requires $w' > \frac{1}{\Lambda}m w_E + \frac{1}{\Lambda} w_R$. Since our choice of $w'$ guarantees $w' \geq (\Lambda m-(\Lambda+1)t) w_E + \Lambda w_R$, it would be enough to ensure $\Lambda m-(\Lambda+1)t \geq \frac{1}{\Lambda}m$  for this (using the fact that $\Lambda > 1$). However, this only holds if $t \leq \frac{\Lambda-1}{\Lambda} m$. To generalize our technique for larger $t$ values, we can add further pairs of fixed nodes (i.e., `fake enablers' as seen earlier) with weight $w_E$, until the new number of enablers $m'$ satisfies $t + \frac{m'-m}{2} \leq \frac{\Lambda-1}{\Lambda} m'$.

\paragraph*{Uses of enabler sets}

This also allows us to devise constructions where we can enable and disable a follower node multiple (constantly many) times throughout a given period. Assume that we add $m$ new enablers to a follower node $v$, and select the weight of its dominant node such that we need $t:=\frac{m}{2}$ out of these nodes besides $v_d$ to dominate $v$. We then use some of these new nodes as actual enabler nodes, and turn the rest into fixed nodes. Note that for the behavior of $v$, there are in fact 3 different cases. If exactly $t$ of the new nodes are black, $v$ behaves as if it was a follower of $v_d$. If less than $t$ are black, then $v_d$ semi-dominates $v$ for white, but $v_d$ being black does not make a black $v$ switchable. Finally, if more than $t$ are black, then $v_d$ semi-dominates $v$ for black.

Thus, using enabler sets essentially allows us to alternate between these 3 behavior patterns. By switching the enablers in the correct order throughout the procedure, we can change arbitrarily between the 3 behaviors, effectively enabling and disabling (\textit{locking}) $v$ multiple times before the enablers are reset. Note that in case of enabling steps, the benevolent player is automatically forced to switch the enabler in question in order to continue execution. However, in case of locking, we need a further check with a logical gate to ensure that the locking step has indeed been completed; otherwise, the benevolent player could avoid locking the enabler set, and he could continue execution with an undesired sequence.

\paragraph*{Enablers with predefined weights}

There is one last case we need to consider. In the previous settings, we always had the freedom to select an arbitrary weight value for the newly added enabler node. However, occasionally, there already is a set of neighbors of a node $v$ with predefined weight values, and we need to use these specific nodes as enablers for $v$.

Assume we need to use enabler nodes $v_{e_1}, ..., v_{e_m}$ to $v$ such that each of the enablers already has a predetermined weight, and for some threshold $w_t$, we want to enable $v$ when the enablers of the same color have a weight of at least $w_t$ together. In this setting, similarly as before, we can select $w'$ such that $v$ is enabled exactly in the desired case. The difference in this case is that, as the weights $w(v_{e_i})$ are already given, we cannot ensure that $w(v_{e_i}) > w_R$. Thus we will only assume this setting within gadgets, where the operation of the gadget ensures that $v_d$ and the nodes composing $w_R$ can not be the same color as $v$ at the same time, and thus it does not matter if $w'+w_R$ is enough to dominate $v$. 

\paragraph*{Observer nodes}

In this setting, our aim is to add a new neighbor $v_o$ to a node $v$ such that the behavior of $v$ is unaffected by this extra neighbor: for each $S \subseteq N(v)$, both $S$ and $S \cup \{ v_o \}$ dominate $v$ after the addition exactly if $S$ dominated $v$ before. Using the same method as with enabler nodes is not straightforward here, as $v$ may have multiple neighbors that have to be updated after the addition of $v_o$. Instead, we show another technique to add observer nodes.

For the desired behavior, we naturally want to assign a very small weight to $v_o$. However, if we have a set $S \subseteq N(v)$ such that $W_S= \frac{\Lambda}{\Lambda+1} \cdot W_{N(v)}$ exactly, then no weight is sufficiently small, since with any positive choice of $w(v_o)$, $S \cup \{ v_o \}$ will dominate $v$, but $S$ alone will not.

As there are only finitely many subsets $S$ of $N(v)$, we can define $w_{\text{next}}(v) = \text{max} \, (W_S \: | \: W_S < \frac{\Lambda}{\Lambda+1} \cdot W_{N(v)} )$, and then define $w_0(v)$ as an arbitrary number between $w_{\text{next}}(v)$ and $\frac{\Lambda}{\Lambda+1} \cdot W_{N(v)}$. If the domination threshold for $v$ was $w_0(v)$, then it would be possible to select a small enough $w(v_o)$ value, since no subset of $N(v)$ has a weight of exactly $w_0(v)$. To make $w_0(v)$ the domination threshold, we could consider Rule II not with the current $\Lambda$ parameter, but with $\Lambda'(v)= \frac{w_0(v)}{W_{N(v)}-w_0(v)} $ instead (i.e., the choice of $\Lambda'(v)$ that ensures $w_0(v) = \frac{\Lambda'(v)}{\Lambda'(v)+1} \cdot W_{N(v)}$).

This suggest the following solution: we first assign weights to every other node in the graph, and consider the observer nodes last. We then consider the $\Lambda'(v)$ value for each node in the graph, and define $\Lambda' = \text{max}_{v \in V} \, \Lambda'(v)$. We now consider the same construction under Rule II with parameter $\Lambda'$. The behavior of the construction under this new rule is exactly the same, since $\Lambda'$ was chosen such that each node is dominated by exactly the same subsets of its neighbors as before; however, we can now select appropriately small weights for observer nodes, giving us a valid benevolent construction for $\Lambda'$. Since we have an adversarial construction for any $\Lambda$ value, we can obtain a benevolent construction for a specific $\Lambda'$ value by simply starting from a construction with $\Lambda > \Lambda'$.

Thus this technique allows us to add observer node neighbors with the desired properties. Note that by choosing $w(v_o)$ sufficiently small, we can also add multiple observer nodes to the same $v$. Also, by selecting the minimum of multiple weights, we can connect the same observer node to multiple nodes in the graph.

\paragraph*{Observer nodes in gates}

Since our technique requires that observer nodes are the last nodes in the graph to be assigned weights, the weight of no other node can depend on the weight of observers. In our construction, we only use observer nodes in logical gates, where $w(v_o)$ only influences the behavior of two specific nodes in the gate (see the description of logical gates for more detail). We now discuss how to select weights to these dependent nodes without knowing $w(v_o)$ in advance, such that the correct behavior of the gadget is ensured in the end.

One of the dependent nodes, $v_A$, is a follower node, and has $v_o$ as a lower neighbor. Here, our job is to ensure that the upper neighbor of $v_A$ will indeed dominate $v_A$, regardless of our choice of $w(v_o)$ later. Since $w(v_o)$ can be arbitrarily small, this can be achieved by defining an artificial upper bound on $w(v_o)$ beforehand, assigning weight to $v_A$'s neighbors based on this bound, and decreasing $w(v_o)$ to this bound in case it would receive a larger value by default.

The other node $v_r$ needs to use $v_o$ as an enabler. For $v_r$, we can select for its dominant node $v_B$ a weight of $w(v_B) = \frac{\Lambda}{\Lambda +1} \cdot W_{N(v_r)}$, allowing us to assign weight to other nodes dependent on $w(v_B)$. Then when the value of $\Lambda'$ is obtained, we add fixed node neighbors to $v_r$ of both colors with weight $w_f$ such that $w(v_B) + w_f = \frac{\Lambda'}{\Lambda' +1} \cdot ( W_{N(v_r)} + 2 w_f)$ (as in the proof of benevolent monotonicity in Appendix \ref{App:C}). This way, $w(v_B)$ will exactly be on threshold value of dominating $v_r$, so $v_o$ will act as an enabler node for any positive choice of $w(v_o)$.

\paragraph*{Pacer systems}

The general idea of pacers has already been shown in Figure \ref{fig:pacer_base}: by adding two relay nodes between a 6-tuple of control nodes and a 6-tuple of base nodes, we obtain a new configuration where base nodes still behave as before, but we can use the newly added pacer nodes to recharge some of our gadgets. Note that the choice of adding 2 pacers in the figure is only for convenience; the technique also works with a chain of any even number of pacer nodes if each of the pacers is a follower of its upper neighbor and the chain has the appropriate alternating color pattern initially. Generally, we say that a node is a pacer node in level $\ell$ if it switches its color exactly once between each consecutive pair of switches of the base nodes in level $\ell$.

Let us now consider a more general setting. Assume we have a set of gadgets that each require one (or more) pacer nodes with a certain weight and certain initial color in level $\ell$. We show how to create a pacer system that ensures the correct behavior of all these pacer nodes. Furthermore, in each gadget, switching the corresponding pacer node forces some color changes within the gadget (i.e, resets it to its default state). Therefore, in the benevolent model, we also need to ensure that the player executes these changes within the gadget after the pacer node is switched. In each of our gadgets, we can select one (or more) specific \textit{signal nodes} such that the resetting of the gadget is finished when the signal node switches to a certain color. Thus for the general pacer setting, we assume that there is also a set of signal nodes with a given weight and a given signal color, and pacer system has to ensure that the player can only continue execution once all these signal nodes are set to the required color.

The general design of such a pacer system is illustrated in Figure \ref{fig:pacersys}. Assume that we place the pacer system above a black 6-tuple of base nodes; this implies that the topmost node in the pacer system (the new lower neighbor of the control nodes in the level) must also be black initially. Each gadget that requires an initially white pacer node can simply have its pacer directly connected to this topmost node as a lower neighbor. Gadgets requiring initially black pacers must first be connected to a white relay node, and only this relay node is connected to the topmost node. We can choose the weight of each such relay node and the topmost node sufficiently high to ensure that each of these nodes is a follower of its upper neighbor. This way, whenever the topmost node switches, it eventually makes each of the pacers in the gadgets switchable.

We need a couple of more pacer nodes in the lower part of the system for checking the conditions on signal nodes. Note that we may expect different behavior in the gadgets in the two possible switching directions in the system, i.e. when the topmost node switches white and thus each pacer switches to the opposite of its initial color (the \textit{pace-away} phase), and when the topmost node switches to black and each pacer switches back to its original color (\textit{pace-back} phase). For both of these directions, we insert a pair of pacers that check the required conditions on the signal nodes: one that has each signal node that needs to be black as an asymmetric enabler, and one that has each signal node that needs to be white as an asymmetric enabler. As seen before, this can always be done with an appropriate choice of weight for the upper neighbor of the node. Note that it is possible for a signal node to be connected to more than one of these pacer nodes, and thus participate in the continuation condition in both switching directions.

This gives us a gadget with the desired properties. Whenever the control nodes in the level switch, they are followed by the topmost pacer node, and then all other pacers eventually. Once execution finishes in the connected gadgets and all signal nodes switch to the required color, the two condition-checking pacers at the bottom switch. The other pair of checking pacers (corresponding to the other switching direction) simply act as follower nodes in this case. Eventually, the lowermost pacer node in the system switches, forcing the base nodes in the level to switch.

Note that in our recursive procedure, between two consecutive times of passing execution down to lower levels, the base nodes always switch an even number of times. This implies that whenever execution is on a lower level, the pacer nodes in the current level always have the same color. Throughout the operation of the gadgets, these pacers behave as if they were fixed nodes, and they only switch in the process of resetting when execution has been passed back to the current level.

\begin{figure}
\centering
\minipage{0.45\textwidth}
	\centering
	\scalebox{.85}{\definecolor{dgray}{gray}{0.25}

\begin{tikzpicture}

	\draw (0pt,0pt) -- (0pt,115pt);
	
	\draw[dashed] (0pt,0pt) -- (-20pt,-14pt);
	\draw[dashed] (0pt,0pt) -- (-12pt,-14pt);
	\draw[dashed] (0pt,0pt) -- (-4pt,-14pt);
	\draw[dashed] (0pt,0pt) -- (4pt,-14pt);
	\draw[dashed] (0pt,0pt) -- (12pt,-14pt);
	\draw[dashed] (0pt,0pt) -- (20pt,-14pt);
	
	\draw[dashed] (0pt,115pt) -- (-20pt,129pt);
	\draw[dashed] (0pt,115pt) -- (-12pt,129pt);
	\draw[dashed] (0pt,115pt) -- (-4pt,129pt);
	\draw[dashed] (0pt,115pt) -- (4pt,129pt);
	\draw[dashed] (0pt,115pt) -- (12pt,129pt);
	\draw[dashed] (0pt,115pt) -- (20pt,129pt);
	
	\draw (-60pt,60pt) to[out=275,in=170] (0pt,0pt);
	\draw (-45pt,60pt) to[out=280,in=165] (0pt,12pt);
	\draw (-30pt,60pt) to[out=275,in=145] (0pt,12pt);
	\draw (-30pt,60pt) to[out=305,in=175] (0pt,42pt);
	
	\draw (-30pt,85pt) -- (-30pt,97pt);
	\draw (-45pt,85pt) -- (-45pt,97pt);
	
	\draw (-30pt,97pt) to[out=55,in=195] (0pt,115pt);
	\draw (-45pt,97pt) to[out=60,in=192] (0pt,115pt);
	\draw (-60pt,85pt) to[out=90,in=185] (0pt,115pt);
	\draw (-75pt,85pt) to[out=80,in=180] (0pt,115pt);

	\draw[black, fill=white] (0pt,0pt) circle (0.8ex);	
	\draw[fill=black] (0pt,12pt) circle (0.8ex);
	
	\draw[black, fill=white] (0pt,30pt) circle (0.8ex);	
	\draw[fill=black] (0pt,42pt) circle (0.8ex);
	
	\draw[black, fill=white] (-30pt,60pt) circle (0.8ex);
	\draw[fill=black] (-45pt,60pt) circle (0.8ex);
	\draw[black, fill=white] (-60pt,60pt) circle (0.8ex);
	
	\draw[fill=black] (-30pt,85pt) circle (0.8ex);
	\draw[fill=black] (-45pt,85pt) circle (0.8ex);
	\draw[black, fill=white] (-60pt,85pt) circle (0.8ex);
	\draw[black, fill=white] (-75pt,85pt) circle (0.8ex);
	\draw[black, fill=white] (-30pt,97pt) circle (0.8ex);
	\draw[black, fill=white] (-45pt,97pt) circle (0.8ex);
	
	\draw[black, fill=white] (0pt,97pt) circle (0.8ex);
	\draw[fill=black] (0pt,115pt) circle (0.8ex);
	
	\draw[thick, gray, densely dotted] (-23pt,91pt) -- (-82pt,91pt) -- (-82pt,79pt) -- (-23pt,79pt) -- cycle;
	\draw[gray] (-85pt,85pt) -- (-82pt,85pt);
	\node[anchor=east] at (-87pt,88.5pt) {\scriptsize \textit{pacers needed}};
	\node[anchor=east] at (-87pt,81.5pt) {\scriptsize \textit{in gadgets}};
	
	\draw[thick, gray, densely dotted] (-23pt,66pt) -- (-67pt,66pt) -- (-67pt,54pt) -- (-23pt,54pt) -- cycle;
	\draw[gray] (-70pt,60pt) -- (-67pt,60pt);
	\node[anchor=east] at (-72pt,63.5pt) {\scriptsize \textit{signal nodes}};
	\node[anchor=east] at (-72pt,56.5pt) {\scriptsize \textit{in gadgets}};
	
	\draw[thick, gray, densely dotted] (-6pt,49pt) -- (-6pt,23pt) -- (6pt,23pt) -- (6pt,49pt) -- cycle;
	\draw[gray] (6pt,36pt) -- (9pt,36pt);
	\node[anchor=west] at (11pt,39.5pt) {\scriptsize \textit{checkers for}};
	\node[anchor=west] at (11pt,32.5pt) {\scriptsize \textit{pace-away}};
	
	\draw[thick, gray, densely dotted] (-6pt,19pt) -- (-6pt,-7pt) -- (6pt,-7pt) -- (6pt,19pt) -- cycle;
	\draw[gray] (6pt,6pt) -- (9pt,6pt);
	\node[anchor=west] at (11pt,9.5pt) {\scriptsize \textit{checkers for}};
	\node[anchor=west] at (11pt,2.5pt) {\scriptsize \textit{pace-back}};
	
	
\end{tikzpicture}}
	\vspace{1pt}
	\caption{General pacer system design.}
	\label{fig:pacersys}
\endminipage\hfill
\hspace{0.01\textwidth}
\minipage{0.45\textwidth}
	\centering
	\vspace{42pt}
	\scalebox{.85}{\definecolor{dgray}{gray}{0.25}

\begin{tikzpicture}

	\draw (0pt,0pt) -- (30pt,0pt);
	\draw (0pt,0pt) -- (0pt,20pt);
	\draw (15pt,0pt) -- (15pt,20pt);
	\draw (30pt,0pt) -- (30pt,20pt);
	
	\draw[dashed] (0pt,0pt) -- (-20pt,12pt);
	\draw[dashed] (0pt,0pt) -- (-20pt,0pt);
	\draw[dashed] (0pt,0pt) -- (-20pt,-12pt);
	\draw[dashed] (30pt,0pt) -- (50pt,0pt);

	\draw[fill=black] (0pt,0pt) circle (0.8ex);	
	\draw[black, fill=white] (15pt,0pt) circle (0.8ex);
	\draw[fill=black] (30pt,0pt) circle (0.8ex);
	
	\draw[black, fill=white] (0pt,20pt) circle (0.8ex);	
	\draw[fill=black] (15pt,20pt) circle (0.8ex);
	\draw[black, fill=white] (30pt,20pt) circle (0.8ex);
	
	\draw[fill=black] (-20pt,12pt) circle (0.6ex);	
	\draw[black, fill=white] (-20pt,0pt) circle (0.6ex);
	\draw[fill=black] (-20pt,-12pt) circle (0.6ex);
	\draw[black, fill=white] (50pt,0pt) circle (0.6ex);
	
	\node[anchor=south] at (75pt,5pt) {\scriptsize  \textit{pace-back}};
	\draw[very thick, dgray, arrows=-stealth] (60pt,5pt) -- (90pt,5pt);
	
	\draw (120pt,0pt) -- (150pt,0pt);
	\draw (120pt,0pt) -- (120pt,20pt);
	\draw (135pt,0pt) -- (135pt,20pt);
	\draw (150pt,0pt) -- (150pt,20pt);
	
	\draw[dashed] (120pt,0pt) -- (100pt,12pt);
	\draw[dashed] (120pt,0pt) -- (100pt,0pt);
	\draw[dashed] (120pt,0pt) -- (100pt,-12pt);
	\draw[dashed] (150pt,0pt) -- (170pt,0pt);

	\draw[fill=black] (120pt,0pt) circle (0.8ex);	
	\draw[black, fill=white] (135pt,0pt) circle (0.8ex);
	\draw[fill=black] (150pt,0pt) circle (0.8ex);
	
	\draw[fill=black] (120pt,20pt) circle (0.8ex);	
	\draw[black, fill=white] (135pt,20pt) circle (0.8ex);
	\draw[fill=black] (150pt,20pt) circle (0.8ex);
	
	\draw[fill=black] (100pt,12pt) circle (0.6ex);	
	\draw[black, fill=white] (100pt,0pt) circle (0.6ex);
	\draw[fill=black] (100pt,-12pt) circle (0.6ex);
	\draw[black, fill=white] (170pt,0pt) circle (0.6ex);
	
	\node[anchor=south] at (75pt,38pt) {\scriptsize  \textit{resetting – pace-away}};
	\draw[very thick, dgray, arrows=-stealth] (135pt,53pt) -- (135pt,39pt) -- (15pt,39pt) -- (15pt,29pt);
	
	\draw (0pt,70pt) -- (30pt,70pt);
	\draw (0pt,70pt) -- (0pt,90pt);
	\draw (15pt,70pt) -- (15pt,90pt);
	\draw (30pt,70pt) -- (30pt,90pt);
	
	\draw[dashed] (0pt,70pt) -- (-20pt,82pt);
	\draw[dashed] (0pt,70pt) -- (-20pt,70pt);
	\draw[dashed] (0pt,70pt) -- (-20pt,58pt);
	\draw[dashed] (30pt,70pt) -- (50pt,70pt);

	\draw[fill=black] (0pt,70pt) circle (0.8ex);	
	\draw[black, fill=white] (15pt,70pt) circle (0.8ex);
	\draw[fill=black] (30pt,70pt) circle (0.8ex);
	
	\draw[fill=black] (0pt,90pt) circle (0.8ex);	
	\draw[black, fill=white] (15pt,90pt) circle (0.8ex);
	\draw[fill=black] (30pt,90pt) circle (0.8ex);
	
	\draw[fill=black] (-20pt,82pt) circle (0.6ex);	
	\draw[fill=black] (-20pt,70pt) circle (0.6ex);
	\draw[black, fill=white] (-20pt,58pt) circle (0.6ex);
	\draw[black, fill=white] (50pt,70pt) circle (0.6ex);
	
	\node[anchor=south] at (75pt,75pt) {\scriptsize  \textit{activated}};
	\draw[very thick, dgray, arrows=-stealth] (60pt,75pt) -- (90pt,75pt);
	
	\draw (120pt,70pt) -- (150pt,70pt);
	\draw (120pt,70pt) -- (120pt,90pt);
	\draw (135pt,70pt) -- (135pt,90pt);
	\draw (150pt,70pt) -- (150pt,90pt);
	
	\draw[dashed] (120pt,70pt) -- (100pt,82pt);
	\draw[dashed] (120pt,70pt) -- (100pt,70pt);
	\draw[dashed] (120pt,70pt) -- (100pt,58pt);
	\draw[dashed] (150pt,70pt) -- (170pt,70pt);

	\draw[black, fill=white] (120pt,70pt) circle (0.8ex);	
	\draw[fill=black] (135pt,70pt) circle (0.8ex);
	\draw[black, fill=white] (150pt,70pt) circle (0.8ex);
	
	\draw[fill=black] (120pt,90pt) circle (0.8ex);	
	\draw[black, fill=white] (135pt,90pt) circle (0.8ex);
	\draw[fill=black] (150pt,90pt) circle (0.8ex);
	
	\draw[fill=black] (100pt,82pt) circle (0.6ex);	
	\draw[fill=black] (100pt,70pt) circle (0.6ex);
	\draw[fill=black] (100pt,58pt) circle (0.6ex);
	\draw[fill=black] (170pt,70pt) circle (0.6ex);
	
\end{tikzpicture}}
	\vspace{12pt}
	\caption{Phases of using and resetting a logical gate.}
	\label{fig:and_phases}
\endminipage\hfill
\hspace{0.9999\textwidth}
\end{figure}

\paragraph*{Logical gates.}

The basic construction of an \textsc{and} gate has already been shown in Figure \ref{fig:and}. Let us denote the observer node by $v_o$, the enabler node by $v_e$, the relay node between the two by $v_r$, and let us denote the three pacer nodes above them by $v_A$, $v_B$ and $v_C$ from left to right. Assume that $v_o$ observes a set of input nodes $U_{0}$, and $v_e$ enables an output node $u_1$. Note that in the process of resetting the gadget, $v_e$ and all nodes in the upper row also act as signal nodes, so they are also connected to some further nodes in a pacer system.

Assume that the weight of $v_e$ (and the weight of the neighbors outside of the gadget) is already given. We first select $w(v_r)$ and $w(v_C)$ such that $v_C$ is a dominant node and $w(v_r)$ is an enabler of $v_e$. We then select $w(v_B)$ such that it dominates $v_r$ with $v_o$ as an enabler (see further notes in the discussion of observer nodes). Finally, $v_A$ is chosen such that it dominates $v_o$, and the nodes in $U_0$ together form an asymmetric enabler for black. 

We demonstrate the different phases of using a logical gate in Figure \ref{fig:and_phases}. When the condition is fulfilled, then $v_o$, $v_r$ and $v_e$ switch in this order, leading to the 'used' state of the gate. Next time when the pacer system connected to the gate switches, each of the three pacer nodes in the gate switch to the other color. Our choice of weights guarantees that this is enough to force $v_o$ to switch back to black, which then results in $v_r$ and $v_e$ switching back to their original color, too. The fact that $v_e$ is a signal node implies that the pacer system only allows execution to continue if $v_e$ has indeed switched back to black. Finally, in the pace-back phase, each of the upper nodes switches back (this is ensured as they are signal nodes), and the gate returns to its initial phase.

In order to ensure that the gate does not immediately start operating again after this last step, we implicitly assume that the condition checked by the gate is false when resetting happens. Since most of our gates have a certain state of the state chain as one of their inputs, and the pacer system is only active when execution is on the level above the gadget, this is straightforward for most gates, as only a few states correspond to execution being on a higher level. For the remaining gates where the input state allows execution to be above, one can check that there is always another conditions that does not hold while resetting.

Note that it only takes minor modifications to apply the same technique to a range of similar settings. By setting the enabler threshold to the minimum weight of nodes in $U_0$ instead of their sum, we obtain a similar construction for a logical \textsc{or} gate. By inserting not one, but two internal relay nodes into the gadget, we obtain gates for cases when the observer an the enabler nodes are required to have the opposite color.

Furthermore, we can also create gates that consist of multiple layers, that is, use the enabler nodes of some gates as the input for a second layer of gates. This allows us to check more complex conditions (e.g., the \textsc{and} of \textsc{or} conditions), or also allows us to create \textsc{and} conditions for a set of nodes that do not all have the same desired color. Note, however, that since resetting requires the condition of the logical gate to be false at the time, we need to reset such multi-layer gates in multiple steps to ensure that the enabler gates of the first layer are already set back to their initial color before we begin resetting the second layer of gates. This can be solved by using multiple pacer systems below each other (within a level), and using the lower-placed systems to reset the gates in the second and further layers.

\paragraph*{State chain design}

A state chain can be implemented as a chain of alternating-colored nodes, each following its upper neighbor, with a single conflict added to the chain that propagates down step by step (as in Figure \ref{fig:chainstates}). Every possible position of the conflict in the chain corresponds to a different state of the procedure. Note we use the word `state' here in a more abstract sense to refer to a phase of the recursive procedure, or the encoding of this phase in the state chain. However, these phases do not actually correspond to concrete states of the minority process, but instead, to states or sets of states of the minority process on the given level only.

In contrast to storage chains in the adversarial case, the state chain has enabler nodes connected to each of its states to ensure that the player can only proceed to the next state when the steps of the current state are all executed. Propagating down the conflict in the state chain is then a prerequisite condition for the steps of the next state, so in order to follow the procedure, the player has to execute the given steps in the construction and proceed though the states in the chain simultaneously.

\begin{figure}
\centering
	\resizebox{1.0\textwidth}{!}{\definecolor{dgray}{gray}{0.25}

\begin{tikzpicture}

	\draw (0pt,0pt) -- (0pt,120pt);

	\draw[black, fill=white] (0pt,120pt) circle (0.9ex);
	\draw[black, fill=white] (0pt,100pt) circle (0.9ex);
	\draw[black, fill=white] (0pt,80pt) circle (0.9ex);
	\draw[fill=black] (0pt,60pt) circle (0.9ex);
	\draw[fill=black] (0pt,40pt) circle (0.9ex);
	\draw[black, fill=white] (0pt,20pt) circle (0.9ex);
	\draw[fill=black] (0pt,0pt) circle (0.9ex);
	
	\draw[thick, dgray, densely dotted] (8pt,60pt) -- (20pt,60pt) -- (20pt,40pt);
	\draw[thick, dgray, densely dotted] (15pt,132pt) -- (20pt,132pt) -- (20pt,60pt);
	\draw[thick, dgray, densely dotted, arrows=-stealth] (8pt,40pt) -- (33pt,40pt);
	
	\draw (40pt,0pt) -- (40pt,120pt);

	\draw[black, fill=white] (40pt,120pt) circle (0.9ex);
	\draw[black, fill=white] (40pt,100pt) circle (0.9ex);
	\draw[black, fill=white] (40pt,80pt) circle (0.9ex);
	\draw[fill=black] (40pt,60pt) circle (0.9ex);
	\draw[black, fill=white] (40pt,40pt) circle (0.9ex);
	\draw[black, fill=white] (40pt,20pt) circle (0.9ex);
	\draw[fill=black] (40pt,0pt) circle (0.9ex);
	
	\draw[thick, dgray, densely dotted] (48pt,40pt) -- (60pt,40pt) -- (60pt,20pt);
	\draw[thick, dgray, densely dotted] (55pt,132pt) -- (60pt,132pt) -- (60pt,40pt);
	\draw[thick, dgray, densely dotted, arrows=-stealth] (48pt,20pt) -- (73pt,20pt);
	
	\draw (80pt,0pt) -- (80pt,120pt);

	\draw[black, fill=white] (80pt,120pt) circle (0.9ex);
	\draw[black, fill=white] (80pt,100pt) circle (0.9ex);
	\draw[black, fill=white] (80pt,80pt) circle (0.9ex);
	\draw[fill=black] (80pt,60pt) circle (0.9ex);
	\draw[black, fill=white] (80pt,40pt) circle (0.9ex);
	\draw[fill=black] (80pt,20pt) circle (0.9ex);
	\draw[fill=black] (80pt,0pt) circle (0.9ex);
	
	\draw[thick, dgray, densely dotted] (88pt,20pt) -- (96pt,20pt) -- (96pt,10pt);
	\draw[thick, dgray, densely dotted] (88pt,0pt) -- (96pt,0pt) -- (96pt,10pt);
	\draw[thick, dgray, densely dotted, arrows=-stealth] (96pt,10pt) -- (104pt,10pt) -- (104pt,80pt) -- (113pt,80pt);
	
	\draw (120pt,0pt) -- (120pt,120pt);

	\draw[black, fill=white] (120pt,120pt) circle (0.9ex);
	\draw[black, fill=white] (120pt,100pt) circle (0.9ex);
	\draw[fill=black] (120pt,80pt) circle (0.9ex);
	\draw[fill=black] (120pt,60pt) circle (0.9ex);
	\draw[black, fill=white] (120pt,40pt) circle (0.9ex);
	\draw[fill=black] (120pt,20pt) circle (0.9ex);
	\draw[fill=black] (120pt,0pt) circle (0.9ex);
	
	\node[anchor=south] at (150pt,56pt) {\tiny propagate};
	\draw[very thick, dgray, arrows=-stealth] (132pt,56pt) -- (168pt,56pt);
	
	\draw (180pt,0pt) -- (180pt,120pt);

	\draw[black, fill=white] (180pt,120pt) circle (0.9ex);
	\draw[black, fill=white] (180pt,100pt) circle (0.9ex);
	\draw[fill=black] (180pt,80pt) circle (0.9ex);
	\draw[black, fill=white] (180pt,60pt) circle (0.9ex);
	\draw[fill=black] (180pt,40pt) circle (0.9ex);
	\draw[black, fill=white] (180pt,20pt) circle (0.9ex);
	\draw[fill=black] (180pt,0pt) circle (0.9ex);
	
	\draw[thick, dgray, densely dotted] (188pt,80pt) -- (194pt,80pt) -- (194pt,38pt);
	\draw[thick, dgray, densely dotted] (188pt,20pt) -- (194pt,20pt) -- (194pt,38pt);
	\draw[thick, dgray, densely dotted, arrows=-stealth] (194pt,38pt) -- (223pt,38pt);
	\node[anchor=north] at (215pt,39pt) {\tiny lock};
	
	\draw[thick, dgray, densely dotted] (223pt,42pt) -- (200pt,42pt) -- (200pt,100pt);
	\draw[thick, dgray, densely dotted, arrows=-stealth] (200pt,100pt) -- (223pt,100pt);
	\node[anchor=south] at (214.5pt,47pt) {\tiny if};
	\node[anchor=south] at (214.5pt,41pt) {\tiny locked};
	\draw[thick, dgray, densely dotted] (194.75pt,60.5pt) -- (199.75pt,60.5pt);
	
	\draw (230pt,0pt) -- (230pt,120pt);

	\draw[black, fill=white] (230pt,120pt) circle (0.9ex);
	\draw[fill=black] (230pt,100pt) circle (0.9ex);
	\draw[fill=black] (230pt,80pt) circle (0.9ex);
	\draw[black, fill=white] (230pt,60pt) circle (0.9ex);
	\draw[fill=black] (230pt,40pt) circle (0.9ex);
	\draw[black, fill=white] (230pt,20pt) circle (0.9ex);
	\draw[fill=black] (230pt,0pt) circle (0.9ex);
	
	\node[anchor=south] at (250pt,56pt) {\tiny prop.};
	\draw[very thick, dgray, arrows=-stealth] (240pt,56pt) -- (260pt,56pt);
	
	\draw (270pt,0pt) -- (270pt,120pt);

	\draw[black, fill=white] (270pt,120pt) circle (0.9ex);
	\draw[fill=black] (270pt,100pt) circle (0.9ex);
	\draw[black, fill=white] (270pt,80pt) circle (0.9ex);
	\draw[fill=black] (270pt,60pt) circle (0.9ex);
	\draw[fill=black] (270pt,40pt) circle (0.9ex);
	\draw[black, fill=white] (270pt,20pt) circle (0.9ex);
	\draw[fill=black] (270pt,0pt) circle (0.9ex);
	
	\draw[dgray, dashed] (262pt,129pt) -- (278pt,129pt) -- (278pt,71pt) -- (262pt,71pt) -- cycle;
	\draw[dgray, dashed] (278pt,105pt) -- (335pt,90pt);
	\draw[dgray, dashed] (278pt,95pt) -- (335pt,36pt);
	
	\node[anchor=south] at (385pt,115pt) {\footnotesize \textit{at next pacing}};
	
	\draw (350pt,80pt) -- (350pt,40pt);
	\draw[black, fill=white] (350pt,80pt) circle (1ex);
	\draw[fill=black] (350pt,60pt) circle (1ex);
	\draw[black, fill=white] (350pt,40pt) circle (1ex);
	
	\node[anchor=south] at (366.5pt,92.5pt) {\tiny pace-};
	\node[anchor=south] at (368.5pt,87.5pt) {\tiny away};
	\draw[thick, dgray, arrows=-stealth] (356pt,86pt) to[out=40,in=140] (380pt,85pt);
	
	\node[anchor=south, rotate=270] at (389pt,70.5pt) {\tiny prop.};
	\draw[thick, dgray, arrows=-stealth] (390pt,76pt) -- (390pt,65pt);
	
	\draw (385pt,80pt) -- (385pt,40pt);
	\draw[fill=black] (385pt,80pt) circle (1ex);
	\draw[black, fill=white] (385pt,60pt) circle (1ex);
	\draw[black, fill=white] (385pt,40pt) circle (1ex);
	
	\node[anchor=south] at (401.5pt,92.5pt) {\tiny pace-};
	\node[anchor=south] at (403.5pt,88pt) {\tiny back};
	\draw[thick, dgray, arrows=-stealth] (391pt,86pt) to[out=40,in=140] (415pt,85pt);
	
	\draw (420pt,80pt) -- (420pt,40pt);
	\draw[black, fill=white] (420pt,80pt) circle (1ex);
	\draw[black, fill=white] (420pt,60pt) circle (1ex);
	\draw[black, fill=white] (420pt,40pt) circle (1ex);
	
	\draw[thick, dgray, densely dotted, arrows=-stealth] (390pt,60pt) -- (394pt,60pt) -- (394pt, 28pt) -- (407pt, 28pt) -- (407pt, 16pt);
	\draw[thick, dgray, densely dotted, arrows=-stealth] (425pt,80pt) -- (429pt,80pt) -- (429pt, 28pt) -- (416pt, 28pt) -- (416pt, 16pt);
	
	\node[anchor=north] at (411.5pt,19pt) {\tiny \textit{checked in}};
	\node[anchor=north] at (411.5pt,13pt) {\tiny \textit{pacer system}};
	
	\draw[gray] (-5pt,-18pt) -- (-5pt,-21pt) -- (85pt, -21pt) -- (85pt, -18pt);
	\draw[gray] (40pt,-21pt) -- (40pt,-24pt);
	\node[anchor=north] at (40pt,-23pt) {\footnotesize \textit{valid states}};
	
	\draw[gray] (115pt,-18pt) -- (115pt,-21pt) -- (275pt, -21pt) -- (275pt, -18pt);
	\draw[gray] (195pt,-21pt) -- (195pt,-24pt);
	\node[anchor=north] at (195pt,-24pt) {\footnotesize \textit{resetting}};

\end{tikzpicture}}
	\caption{Illustration of a state chain with 3 states. Dotted lines show conditions and enabled nodes of \textsc{AND} gates; lines with no source indicate other conditions outside of the chain. After resetting, the two conflicts stored in the extra nodes are reintroduced in the next pacing.}
	\label{fig:statechain}
\end{figure}
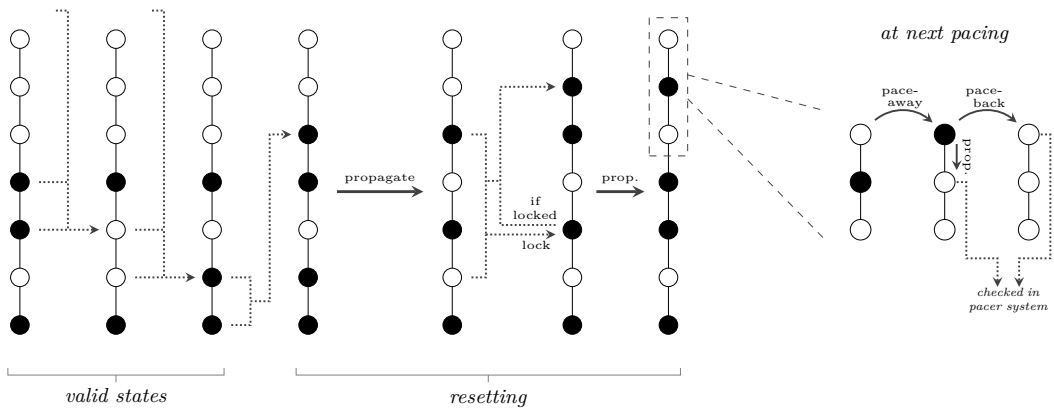

In order to reset the state chain, we add 3 extra nodes and store 2 more conflicts at the top of the chain. Whenever the lowermost state is reached, we let the first stored conflict propagate through the chain, resetting each node to its initial color. We then allow the second stored conflict to propagate down to the position that corresponds to the first state, resetting the chain to its original state (except for the uppermost extra nodes). We include the uppermost node in the chain as a pacer in the pacer system in the level above, and use the pacer system to ensure that this uppermost node always switches the second node while pacing. This ensures that when execution is passed back to the level above, the two required conflicts are reintroduced into the chain. The behavior of the chain is illustrated in Figure \ref{fig:statechain}.

Note that besides enabling some nodes, we occasionally also have to lock nodes in the process to ensure that the player can only follow the desired sequence of steps in the state chain. One such locking is specifically shown in the figure.

\paragraph*{States of the procedure}

We now describe the states needed in the state chain in order to encode the procedure that the benevolent player has to follow. As mentioned, to follow the recursive procedure throughout the various levels, we need specific states in the state chain of each level to indicate if execution is currently on a level above or below. If a level needs to pass execution to a neighboring level (say, the one below), then we proceed to the next, `level below' state into the current level's state chain. Also, in the state chain of the level below (which was currently in a `level above' state), we proceed to the next state as well, which encodes a state in which the execution is indeed happening on this lower level.

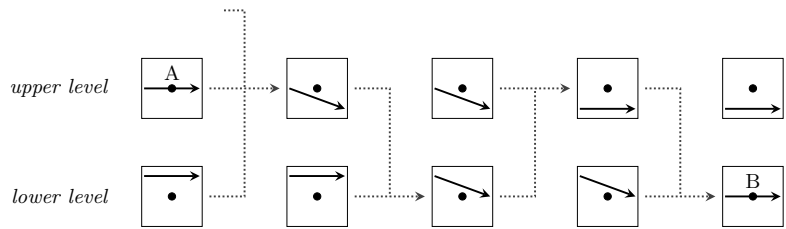
\begin{figure}
\centering
	\resizebox{0.75\textwidth}{!}{\definecolor{dgray}{gray}{0.25}

\begin{tikzpicture}

	\node[anchor=east] at (-10pt,12.5pt) {\footnotesize \textit{lower level}};
	\node[anchor=east] at (-10pt,57.5pt) {\footnotesize \textit{upper level}};

	\draw (0pt,0pt) -- (0pt,25pt) -- (25pt, 25pt) -- (25pt, 0pt) -- cycle;
	\draw[fill=black] (12.5pt,12.5pt) circle (0.35ex);
	\draw[thick, arrows=-stealth] (1pt,21pt) -- (24pt,21pt);
	
	\draw (0pt,45pt) -- (0pt,70pt) -- (25pt, 70pt) -- (25pt, 45pt) -- cycle;
	\draw[fill=black] (12.5pt,57.5pt) circle (0.35ex);
	\draw[thick, arrows=-stealth] (1pt,57.5pt) -- (24pt,57.5pt);
	\node[anchor=south] at (12.5pt,57.5pt) {\footnotesize A};
	
	\draw[thick, dgray, densely dotted, arrows=-stealth] (28pt,57.5pt) -- (57pt,57.5pt);
	\draw[thick, dgray, densely dotted] (28pt,12.5pt) -- (42.5pt,12.5pt) -- (42.5pt,57.5pt);
	\draw[thick, dgray, densely dotted] (34pt,90pt) -- (42.5pt,90pt) -- (42.5pt,57.5pt);	
	
	\draw (60pt,0pt) -- (60pt,25pt) -- (85pt, 25pt) -- (85pt, 0pt) -- cycle;
	\draw[fill=black] (72.5pt,12.5pt) circle (0.35ex);
	\draw[thick, arrows=-stealth] (61pt,21pt) -- (84pt,21pt);
	
	\draw (60pt,45pt) -- (60pt,70pt) -- (85pt, 70pt) -- (85pt, 45pt) -- cycle;
	\draw[fill=black] (72.5pt,57.5pt) circle (0.35ex);
	\draw[thick, arrows=-stealth] (61pt,57.5pt) -- (84pt,49pt);
	
	\draw[thick, dgray, densely dotted, arrows=-stealth] (88pt,12.5pt) -- (117pt,12.5pt);
	\draw[thick, dgray, densely dotted] (88pt,57.5pt) -- (102.5pt,57.5pt) -- (102.5pt,12.5pt);
	
	\draw (120pt,0pt) -- (120pt,25pt) -- (145pt, 25pt) -- (145pt, 0pt) -- cycle;
	\draw[fill=black] (132.5pt,12.5pt) circle (0.35ex);
	\draw[thick, arrows=-stealth] (121pt,21pt) -- (144pt,12.5pt);
	
	\draw (120pt,45pt) -- (120pt,70pt) -- (145pt, 70pt) -- (145pt, 45pt) -- cycle;
	\draw[fill=black] (132.5pt,57.5pt) circle (0.35ex);
	\draw[thick, arrows=-stealth] (121pt,57.5pt) -- (144pt,49pt);
	
	\draw[thick, dgray, densely dotted, arrows=-stealth] (148pt,57.5pt) -- (177pt,57.5pt);
	\draw[thick, dgray, densely dotted] (148pt,12.5pt) -- (162.5pt,12.5pt) -- (162.5pt,57.5pt);
	
	\draw (180pt,0pt) -- (180pt,25pt) -- (205pt, 25pt) -- (205pt, 0pt) -- cycle;
	\draw[fill=black] (192.5pt,12.5pt) circle (0.35ex);
	\draw[thick, arrows=-stealth] (181pt,21pt) -- (204pt,12.5pt);
	
	\draw (180pt,45pt) -- (180pt,70pt) -- (205pt, 70pt) -- (205pt, 45pt) -- cycle;
	\draw[fill=black] (192.5pt,57.5pt) circle (0.35ex);
	\draw[thick, arrows=-stealth] (181pt,49pt) -- (204pt,49pt);
	
	\draw[thick, dgray, densely dotted, arrows=-stealth] (208pt,12.5pt) -- (237pt,12.5pt);
	\draw[thick, dgray, densely dotted] (208pt,57.5pt) -- (222.5pt,57.5pt) -- (222.5pt,12.5pt);
	
	\draw (240pt,0pt) -- (240pt,25pt) -- (265pt, 25pt) -- (265pt, 0pt) -- cycle;
	\draw[fill=black] (252.5pt,12.5pt) circle (0.35ex);
	\draw[thick, arrows=-stealth] (241pt,12.5pt) -- (264pt,12.5pt);
	\node[anchor=south] at (252.5pt,12.5pt) {\footnotesize B};
	
	\draw (240pt,45pt) -- (240pt,70pt) -- (265pt, 70pt) -- (265pt, 45pt) -- cycle;
	\draw[fill=black] (252.5pt,57.5pt) circle (0.35ex);
	\draw[thick, arrows=-stealth] (241pt,49pt) -- (264pt,49pt);

\end{tikzpicture}}
	\caption{Passing execution to the level below with auxiliary states. The vertical position of arrows in the states shows if execution is on, above or below the current level. Enabling conditions are shown as dotted lines. $A$ and $B$ denote some specific states on the current level.}
	\label{fig:levelchange}
\end{figure}

Note, however, that depending on which of these two steps is done first, there might be points in time when the state chains indicate that execution is in two levels simultaneously, or in no level at all. To simplify the transition from one level to another, we avoid such situations by introducing auxiliary transition states into the state chains for each such transition. The states and conditions of such a transition process are illustrated in Figure \ref{fig:levelchange}.

Now let us consider the whole behavior of the procedure. For the specific case of $\Lambda_B=5$, we can summarize the execution on a level as the periodic repetition of the following steps:

\vspace{24pt}

\renewcommand{\theenumi}{\arabic{enumi}}
\begin{enumerate}
	\item ($P_1$) We execute the first 4 steps of the control sequence (base nodes switch 4 times).
	\item The storage chains below the base nodes are now charged, so we continue execution below until the chains are unloaded.
	\item ($P_2$) We execute the last 2 steps of the control sequence. Base nodes switch 2 times, storage chains below are half-charged.
	\item The storage chains above are unloaded, so we pass execution to the level above. Execution only returns when the storage chains above are charged again.
	\item ($P'_2$) In the control nodes, we execute only the first 2 steps of the control sequence, since storage chains below are charged by then (base nodes switch only 2 times).
	\item Execution is continued below until the storage chains below are unloaded.
	\item ($P_3$) We execute the last 4 steps of the control sequence; storage chains below are charged.
	\item Execution continues below until the chains are unloaded.
	\item The storage chains above are unloaded, so when execution returns from below, we pass it on directly to the level above.
\end{enumerate}

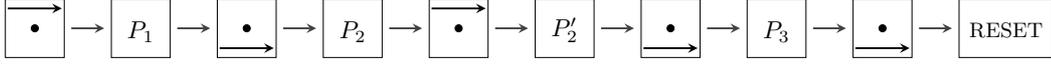
\begin{figure}
\centering
	\resizebox{1.0\textwidth}{!}{\definecolor{dgray}{gray}{0.25}

\begin{tikzpicture}

	\draw (-45pt,0pt) -- (-45pt,25pt) -- (-20pt, 25pt) -- (-20pt, 0pt) -- cycle;
	\draw[fill=black] (-32.5pt,12.5pt) circle (0.35ex);
	\draw[thick, arrows=-stealth] (-44pt,21pt) -- (-21pt,21pt);
	
	\draw[thick, dgray, arrows=-stealth] (-17pt,12.5pt) -- (-3pt,12.5pt);
	
	\draw (0pt,0pt) -- (0pt,25pt) -- (25pt, 25pt) -- (25pt, 0pt) -- cycle;
	\node[anchor=south] at (12.5pt,3pt) {\large $P_1$};
	
	\draw[thick, dgray, arrows=-stealth] (28pt,12.5pt) -- (42pt,12.5pt);
	
	\draw (45pt,0pt) -- (45pt,25pt) -- (70pt, 25pt) -- (70pt, 0pt) -- cycle;
	\draw[fill=black] (57.5pt,12.5pt) circle (0.35ex);
	\draw[thick, arrows=-stealth] (46pt,4pt) -- (69pt,4pt);
	
	\draw[thick, dgray, arrows=-stealth] (73pt,12.5pt) -- (87pt,12.5pt);
	
	\draw (90pt,0pt) -- (90pt,25pt) -- (115pt, 25pt) -- (115pt, 0pt) -- cycle;
	\node[anchor=south] at (102.5pt,3pt) {\large $P_2$};
	
	\draw[thick, dgray, arrows=-stealth] (118pt,12.5pt) -- (132pt,12.5pt);
	
	\draw (135pt,0pt) -- (135pt,25pt) -- (160pt, 25pt) -- (160pt, 0pt) -- cycle;
	\draw[fill=black] (147.5pt,12.5pt) circle (0.35ex);
	\draw[thick, arrows=-stealth] (136pt,21pt) -- (159pt,21pt);
	
	\draw[thick, dgray, arrows=-stealth] (163pt,12.5pt) -- (177pt,12.5pt);
	
	\draw (180pt,0pt) -- (180pt,25pt) -- (205pt, 25pt) -- (205pt, 0pt) -- cycle;
	\node[anchor=south] at (192.5pt,3pt) {\large $P'_2$};
	
	\draw[thick, dgray, arrows=-stealth] (208pt,12.5pt) -- (222pt,12.5pt);
	
	\draw (225pt,0pt) -- (225pt,25pt) -- (250pt, 25pt) -- (250pt, 0pt) -- cycle;
	\draw[fill=black] (237.5pt,12.5pt) circle (0.35ex);
	\draw[thick, arrows=-stealth] (226pt,4pt) -- (249pt,4pt);
	
	\draw[thick, dgray, arrows=-stealth] (253pt,12.5pt) -- (267pt,12.5pt);
	
	\draw (270pt,0pt) -- (270pt,25pt) -- (295pt, 25pt) -- (295pt, 0pt) -- cycle;
	\node[anchor=south] at (282.5pt,3pt) {\large $P_3$};
	
	\draw[thick, dgray, arrows=-stealth] (298pt,12.5pt) -- (312pt,12.5pt);
	
	\draw (315pt,0pt) -- (315pt,25pt) -- (340pt, 25pt) -- (340pt, 0pt) -- cycle;
	\draw[fill=black] (327.5pt,12.5pt) circle (0.35ex);
	\draw[thick, arrows=-stealth] (316pt,4pt) -- (339pt,4pt);
	
	\draw[thick, dgray, arrows=-stealth] (343pt,12.5pt) -- (357pt,12.5pt);
	
	\draw (360pt,0pt) -- (360pt,25pt) -- (400pt, 25pt) -- (400pt, 0pt) -- cycle;
	\node[anchor=south] at (380pt,5pt) {\small RESET};

\end{tikzpicture}}
	\caption{Overview of states in the state chain for $\Lambda_B=5$ (without auxiliary states, for simplicity).}
	\label{fig:statelist}
\end{figure}

Note that with the last step, we arrive back to the initial phase of the process: the storage chains below are completely unloaded, and storage chains above will be fully charged when execution returns next time to this level. This sequence of states is also summarized in Figure \ref{fig:statelist}.

Recall that in a regular step of our benevolent construction, the arrival of the 6 conflicts to a position in the storage chain below acts as the enabling condition for the downpropagation of the next 4-tuple of conflicts in the storage chains above. However, in the different phases, the same conflict positions below must enable different 4-tuples above. In phase $P_1$, the $i^{\text{th}}$ subset in the control sequence inserts the $i^{\text{th}}$ conflict into the storage chain for $i \in \{1, ..., 4\}$. The second phase is split into two parts by the execution being passed above: first in $P_2$, the $(i+4)^{\text{th}}$ control subset produces the $i^{\text{th}}$ conflict into the storage chain for $i \in \{1, 2\}$, and then in $P'_2$, the $i^{\text{th}}$ subset produces the $(i+2)^{\text{nd}}$ conflict for $i \in \{1, 2\}$. Finally, in $P_3$, the $(i+2)^{\text{th}}$ subset of the control sequence creates conflict $i$ in the chain for $i \in \{1, ... 4\}$. Therefore, each logical gate (enabling a 4-tuples of conflicts above) must also include a given state in its condition, and we need to have separate enabler gates for each 4-tuple in the different phases.

One can similarly collect the phases for other $\Lambda_B$ values. For a general (odd) $\Lambda_B$, the list consists of $\frac{\text{LCM}(\Lambda_B-1,\: \Lambda_B+1)}{\Lambda_B-1} = \frac{\Lambda_B+1}{2}$ phases, where $\text{LCM}$ denotes the least common multiple.

\paragraph*{Top of the construction}

Recall that above the uppermost level, we have a constant-length initial storage chain for each of the topmost control nodes, and as such, the uppermost level is irregular in this sense. However, executing the control sequence on this level consists of constantly many steps only, and thus we can hard-code this process into a constant number of logical gates and a constant-size state chain, which do not need to be reset after use.

\section{Generalizations} \label{App:C}

\subsection{Higher number of colors} \label{App:C1}

To generalize our lower bounds for more than two colors, we first have to adjust some of our definitions for this case. Specifically, the different definitions of Rule II are not equivalent anymore: if the sum of black and white neighbors is not fixed, then the difference of two frequencies does not automatically determine their ratio. 

As before, let $c^*$ denote the least frequent color in the neighborhood of a node $v$. Let us introduce $N_{OPT}(v) := \{u \: | \: u \in N(v) \text{ and } C(u)=c^*\}$. We consider the possible definitions of Rule II separately:

\vspace{7pt}

\noindent \textsc{Rule II/}a (for a given $\Lambda$): $v$ is switchable if $W_{N_S(v)} \geq \Lambda \cdot W_{N_{OPT}(v)}$.

\vspace{7pt}

\noindent \textsc{Rule II/}b (for a given $\lambda$): $v$ is switchable if $W_{N_S(v)} - W_{N_{OPT}(v)} \geq \lambda \cdot W_{N(v)}$.

\vspace{7pt}

\noindent \textsc{Rule II/}c (for a given $\rho$): $v$ is switchable if $W_{N_S(v)} \geq \rho \cdot W_{N(v)}$.

\vspace{7pt}

We show how to generalize the bound for all three rules. Given that we already have a construction $G$ on $n$ nodes for 2 colors, we show how to add extra nodes to the graph such that the process behaves as if the graph consisted only of the original nodes with the original two colors. In order to achieve this, for every original node $v$ and each extra color $c \in \Gamma \setminus \{ \textit{black}, \textit{white}\}$, we add an extra neighbor $v_c$ that we only connect to $v$. By selecting the $w(v_c)$ values carefully, we ensure that no original node ever switches to any of the extra colors, and thus the modification does not influence stabilization time. Note that the extra nodes never have any incentive to switch, since they do not have conflicts at all.

First consider Rule II/a. It is clear that selecting $w(v_c)>\frac{1}{2}W_{N(v)}$ is sufficient, since at every point, either the black or white nodes in $N(v)$ have a weight of at most $\frac{1}{2}W_{N(v)}$, so this already ensures that $c$ will never be the preferred color of node $v$. Actually, because of the strict switching condition, choosing only $w(v_c)>\frac{1}{\Lambda+1}W_{N(v)}$ would also suffice.

The case of Rule II/b is slightly more involved. Let $w_0$ denote $\frac{1-\lambda}{2} \cdot W_{N(v)}$, the maximum weight in $N_O(v)$ with which $v$ is still switchable in the original graph $G$. As in the previous case, we again ensure $w(v_c)>w_0$, for example, we select $w(v_c):=\frac{|\Gamma|-1}{|\Gamma|-2} \cdot w_0$. This way, after adding the extra nodes, we have $W'_{N(v)} = W_{N(v)} + |\Gamma-2| \cdot w(v_c) = \left( \frac{2}{1-\lambda} + |\Gamma|-1 \right) \cdot w_0$. The original difference required for switchability was $\lambda \cdot W_{N(v)} = \frac{2 \lambda}{1-\lambda} \cdot w_0 \, $; after adding the extra nodes, this accounts only for a $\frac{\lambda \cdot W_{N(v)}}{W'_{N(v)}} = \frac{2 \lambda}{2 + (|\Gamma|-1)(1 - \lambda)}$ portion of $W'_{N(v)}$. Thus by adding the extra nodes to $G$, we obtain a valid construction for $\lambda':=\frac{2 \lambda}{2 + (|\Gamma|-1)(1 - \lambda)}$ and $|\Gamma|$ colors. Since $\text{lim}_{\lambda \rightarrow 1} \lambda'=1$, this finishes our generalization: for every possible $\lambda'$ value and set of colors $\Gamma$, we can select another (larger) $\lambda$ value such that the 2-color construction with $\lambda$ also provides a construction for $\lambda'$ and $|\Gamma|$.

The case of Rule II/c is almost identical to that of Rule II/b. Now let $w_0 := (1-\rho) \cdot W_{N(v)}$, and again select $w(v_c):=\frac{|\Gamma|-1}{|\Gamma|-2} \cdot w_0$. After the addition of the extra nodes, we get $W'_{N(v)} = W_{N(v)} + |\Gamma-2| \cdot w(v_c) = \left( \frac{1}{1-\rho} + |\Gamma|-1 \right) \cdot w_0$. The original $\rho$ portion of the neighborhood now only accounts for a $\rho':= \frac{\rho \cdot W_{N(v)}}{W'_{N(v)}} = \frac{\rho}{1 + (|\Gamma|-1)(1 - \rho)}$ portion of $W'_{N(v)}$. Again, $\text{lim}_{\rho \rightarrow 1} \rho'=1$, so for any given $\rho'$, we can select a $\rho$ value which indirectly also provides a construction for $\rho'$ and $|\Gamma|$ colors.

Thus we can apply this technique for all three possible generalizations of the proportional switching rule. Assuming only $O(1)$ extra colors, the number of newly added nodes with the technique is $O(n)$, and thus stabilization time remains exponential. Furthermore, in case of Rule II/a, for every extra color $c$, we could also combine the extra nodes $v_c$ (for every $v$) into a single node, selecting the maximum of their weights as the weight of the combined node. This way, we only require one additional node for each newly introduced color, allowing a generalization to up to $\Theta(n)$ colors in case of Rule II/a.

\subsection{Rule II with other $\lambda$ values} \label{App:C2}

In Section \ref{sec:adv}, we have already presented the main idea behind generalizing the adversarial construction to any odd integer $\Lambda_B$. It only remains to discuss the generalization of the control sequence. The description of the control sequence already hints the way of generalizing the method: when starting with $\Lambda_B-1$ white, $1$ black and then $1$ more white node in this order, and continuously "sliding" the ($\Lambda_B-1$)-subset of subsequent nodes to switch (see the dotted line in Figure \ref{fig:combinat}), we can always produce a control sequence of $\Lambda_B+1$ steps where each control node switches $\Lambda_B-1$ times altogether, and the base node below is switched $\Lambda_B+1$ times. Besides the control sequence, there is one more detail to note for completeness: the length of the initial starting chain above the uppermost level also grows for higher $\Lambda_B$, but only to a constant value which is a function of $\Lambda_B$.

As discussed earlier, with an appropriate choice of $\Lambda_B$ and $\epsilon$, this proves the exponential lower bound in model $SA$ for any parameter $\lambda \in (0,1)$. The techniques in Section \ref{sec:ben} were shown for a general $\Lambda$ value in the first place, thus the lower bound also holds for any $\lambda \in (0,1)$ in models $SB$ and $CB$.

To gain a deeper understanding of the process, we also present a simple method to prove monotonicity of the $\lambda$ values: that is, given a two values $\lambda_0$ and $\lambda$ with $\lambda_0 > \lambda$, and given a construction with exponential stabilization time for $\lambda_0$, we can transform this into another construction which has exponential stabilization time for $\lambda$. Recall that this transformation is trivial in the adversarial case, as the same graph with the same sequence of steps is also a valid construction for $\lambda$. In the benevolent case, the idea is to add, for each node $v$ in the graph, two fixed node neighbors to $v$ (a black and a white one) to reduce the original weight $W_{N(v)}$ to stand for only a given portion of the weight in the neighborhood of $v$. There is a specific weight value $w_f$ that we can assign to these two fixed nodes such that the resulting graph behaves the same way under Rule II with $\lambda$ as the original graph behaved with $\lambda_0$. Indeed, since the new nodes can never switch but contribute a weight of $w_f$ to each color, we can achieve this by ensuring $\frac{\Lambda_0}{\Lambda_0+1} \cdot W_{N(v)} + w_f = \frac{\Lambda}{\Lambda+1} \cdot (W_{N(v)}+2 w_f)$. This requires a choice of $w_f=\frac{1}{2 \lambda} \cdot \left( \lambda_0 - \lambda \right) \cdot W_{N(v)}$.

\end{appendices}

\end{document}